\documentclass[12pt]{article}

\usepackage{siunitx}
\usepackage{authblk}
\usepackage{graphicx}%
\usepackage{array}
\usepackage{soul}
\usepackage{multirow}%

\usepackage{amsmath}%
\usepackage{amssymb,amsfonts}
\usepackage{amsthm}%
\usepackage{mathrsfs}%
\usepackage[title]{appendix}%
\usepackage{xcolor}%
\usepackage{textcomp}%
\usepackage{manyfoot}%
\usepackage{booktabs}%
\usepackage{algorithm}%
\usepackage{algorithmicx}%
\usepackage{algpseudocode}%
\usepackage{listings}%
\usepackage{natbib}
\usepackage{threeparttable}
\usepackage{subcaption}
\usepackage{hyperref}
\usepackage[T1]{fontenc}
\usepackage[utf8]{inputenc}  
\usepackage[T1]{fontenc}     
\usepackage{textcomp}        

\makeatletter

\def\uns{\ifmmode\,\else$\,$\fi}%

\makeatother

\title{Inverse Velocity Dispersion of Solar Energetic Protons Observed by Solar Orbiter and Its Shock Acceleration Explanation}

\author[1,3]{Yuncong Li}
\author[1,2]{Jingnan Guo\thanks{Corresponding author: \texttt{jnguo@ustc.edu.cn}}}
\author[1]{Daniel Pacheco\thanks{Corresponding author: \texttt{dpacheco@ustc.edu.cn}}}
\author[1,2]{Yuming Wang\thanks{Corresponding author: \texttt{ymwang@ustc.edu.cn}}}
\author[3]{Manuela Temmer}
\author[4]{Zheyi Ding}
\author[4]{Robert F. Wimmer-Schweingruber}

\affil[1]{National Key Laboratory of Deep Space Exploration / School of Earth and Space Sciences, University of Science and Technology of China, Hefei, China}
\affil[2]{CAS Center for Excellence in Comparative Planetology, University of Science and Technology of China, Hefei, China}
\affil[3]{Institute of Physics, University of Graz, Universitätsplatz 5/2, Graz, Austria}
\affil[4]{Institute of Experimental and Applied Physics, Kiel University, Kiel, Germany}

\date{\today}

\begin{document}
\maketitle

\begin{abstract}
The particle acceleration and transport process during solar eruptions is one of the critical and long-standing problems in space plasma physics. Through decades of research, it is well accepted that particles with higher energies released during a solar eruption arrive at observers earlier than the particles with lower energies, forming a well-known structure in the dynamic energy spectrum called particle velocity dispersion (VD), as frequently observed by space missions. However, this picture is challenged by new observations from NASA’s Parker Solar Probe and ESA’s Solar Orbiter which show an unexpected inverse velocity dispersion (IVD) phenomenon, where particles with higher-energies arrive later at the observer. Facing on the challenge, we here report the recent discovery of such IVD structures with 10 solar energetic proton events observed by Solar Orbiter, and then analyze the mechanisms causing this unusual phenomenon. We suggest that shock diffusive acceleration, with respect to magnetic reconnection, is probably a dominant mechanism to accelerate protons to tens of MeV in such events where particles need longer time to reach higher energies. And we determine, innovatively, the physical conditions and time scales during the actual shock acceleration process that cannot be observed directly. 
\end{abstract}

\noindent\textbf{Keywords:} SEP, CME shocks, Particle Acceleration, Magnetic Connectivity

\section{Introduction}
Solar energetic particles (SEP) are accelerated by flares and/or shocks driven by coronal mass ejections (CMEs) \citep{Reames2013,Klein2017}. SEPs constitute an important radiation hazard for robotic and crewed space missions and their understanding is of capital importance to assure the success of future space explorations \citep{Guo24}. 
Previous observations often showed that more energetic (and therefore faster) particles arrived earlier than less energetic particles during solar eruptions \citep{Laitinen2015,Gomez-Herrero2021,Kollhoff2021}. This supports the idea that {the acceleration time scale is relatively short so that the accelerated} particles are released at approximately the same time irrespective of their energy and propagate along similar trajectories. This phenomenon is called particle velocity dispersion  \citep[VD, e.g.,][]{wimmer2023a,vanio2013,Malandraki2012}. 

Since the launch of Solar Orbiter (SolO) in 2020 \citep{Mueller2020}, it has collected a detailed dataset of remote and in-situ observations of solar eruptions. In particular, the unprecedented resolution of energetic particle observations has opened a new window looking into the fine structures of SEP events and dynamics \citep{wimmer2023a, Aran2021, Lario2022, Yang2024}. 
As solar activity has been rising while entering in the solar maxima of cycle 25, SolO has observed an increasing number of events, some of which defy our previous understanding of VD. 
During these unusual events we observe the known VD feature for particles below an event-specific energy, but particles at higher energies appear to arrive later with increasing energy. This behaviour has been referred as ``inverse velocity dispersion" (IVD) based on a single case observed by Parker Solar Probe (PSP) when it was very close to the Sun \citep[at around 0.07 au,][]{Cohen2024, Kouloumvakos2025}. 
Until the end of 2024, SolO has observed at least 10 of such events, however, at all distances from the Sun, from 0.49 - 0.95 au, and for a wide range of relative longitudinal separations between the solar source and SolO. 

In this paper we present the 10 events observed so far and analyse three especially clear examples in more detail.
{We find that the IVD features can be well interpreted by the diffusive shock acceleration (DSA) mechanism through which particles need longer time to obtain higher energies before being released from the acceleration site. This finding suggests that DSA is probably a dominant mechanism in such events, with respect to magnetic reconnection, to accelerate protons to tens of MeV which could be potentially radiation damaging to instruments and humans in space under thin shielding. We further determine, innovatively, the physical conditions and time scales during the actual acceleration process of these particles at the shock that can not be observed directly. }

\section{Results}\label{sec:1109section}

Since early 2022 to mid-2024, SolO has observed many SEP events among which we select those with IVD features using the following criteria.
\begin{enumerate}
    \item The event must have a clean background, unaffected by the preceding events at the main energy range covered by 
    SolO's Energetic Particle Detector \citep[EPD,][]{RodriguezPacheco2020,Wimmer21}.
    \item The proton increase must show clear VD and IVD parts, with a transition of the two structures from low to high energies.
    \item SolO must be in front of the solar disk as seen from Earth so that Earth-based remote-sensing observations can be useful (see Section Dataset of Supplementary Information (SI)).
    \item White-light coronagraphs onboard the Solar-Terrestrial Relations Observatory Ahead \citep[STEREO A,][]{Kaiser2008} and the Solar and Heliospheric Observatory \citep[SOHO,][]{Domingo95} should observe the CME simultaneously so that the original direction and speed of the CME can be derived (see Section Dataset of SI).
    
\end{enumerate}

A total of 10 IVD events were selected based on the first two criteria, with only 3 events satisfying all four conditions and being analysed in detail.
We first present a detailed analysis of the 2023-11-09 event because it had the clearest association of flare, CME, and SEPs. Together with another two events (2023-12-24 and 2023-12-31), the results are summarized in Tab.~\ref{tab:results}. Further information for all 10 IVD events is given in Tab.~\ref{tab:results2} and SI.

\subsection{Overview of the 2023-11-09 event} 
\begin{figure*}
\resizebox{\textwidth}{!}
    {\includegraphics{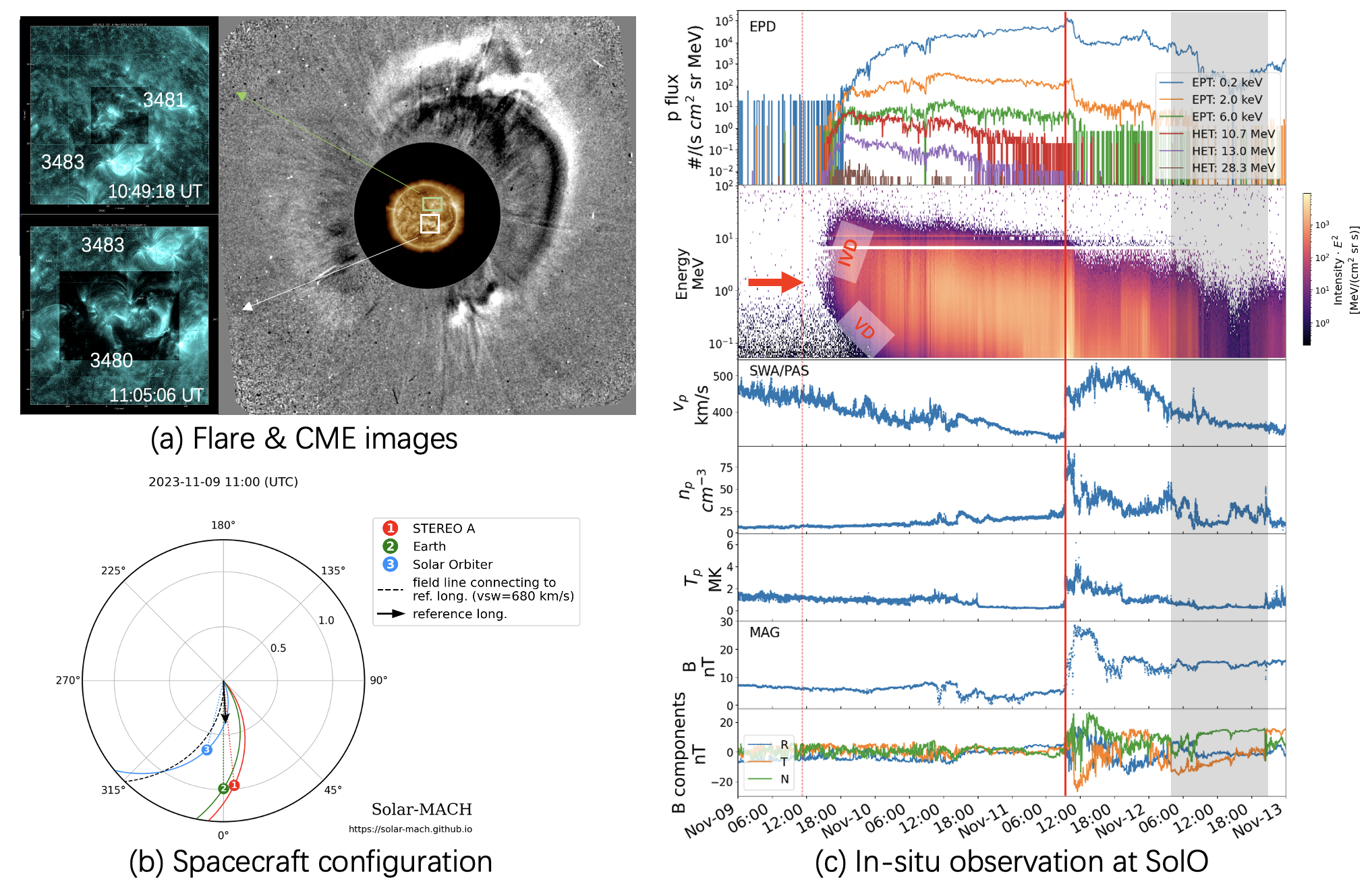}} 
    \caption{Overview of the solar eruption on 2023-11-09 and in-situ observations of the SEP event and interplanetary environment. Panel (a, right) shows the EUV (131 \AA) observation of the Sun and white-light coronagraph observation of the CMEs from Earth's view. Panel(a, left) shows zoomed-in images of the two candidate source regions (flares). (b) shows the flare direction, positions and magnetic connectivity of SolO and other spacecraft in the solar ecliptic plane. (c) shows the in-situ measurements at SolO including, from top to bottom, energetic proton flux for 6 different energy bins (shown in the legend), energetic proton dynamic spectrum across the whole energy range from $\sim$ 50 keV to 105 MeV, solar wind bulk speed, proton density, proton temperature, magnetic field magnitude, the magnetic field vector components in RTN coordinates. The red dotted line marks the peak of the flare shown by the SolO hard X-ray observations (also see SI and Table 1); the red solid line indicates the shock arrival at SolO; the gray-shaded area marks the duration of the CME that drove the shock. All particle measurements were obtained from EPD's Sunward-looking telescopes. }
    \label{fig:1109}
\end{figure*}

Using solar observations (Section: Dataset in SI), the 2023-11-09 SEP event could be associated with two C-class flares (C1.3 observed between 10:41  and 11:07~UT from AR 13481 located at N24W15 and C2.6  observed between 10:53  and 11:37~UT from AR13480 located at S10W04) accompanied by a wide halo CME as seen from Earth (Fig. \ref{fig:1109}a). 
The EUV observation of the Sun (in 193 \AA) shows the two flares and their zoomed-in images (in 131 \AA). The CME shock can be clearly seen in the running difference image of the white-light coronagraph. A movie of this time period showing the flare and CME eruption can be found here (\url{https://cdaw.gsfc.nasa.gov/movie/make\_javamovie.php?date=20231109&img1=sdo\_a193&img2=lasc2rdf}) which indicates that there may be a prior CME heading north that interacted with the halo CME while the latter is most likely related to the C2.6 flare located in the southern hemisphere.
The speed of the halo CME-driven shock is derived to be 870 km/s based on quasi-graduated cylindrical shell  \citep[GCS,][]{temmer2015interplanetary} fits when the shock was clearly shown in the coronagraph.

The CME speed and flare classes were not among the highest on record, nevertheless, clear solar radio bursts were observed which are associated with the shock propagation and particle release process. 
Type \uppercase\expandafter{\romannumeral3} emission was observed at the beginning of the flare eruption and a slow drift of Type \uppercase\expandafter{\romannumeral2} radio burst started from $\sim$1 MHz to 20 kHz (see SI).

Figure~\ref{fig:1109}(b) shows the positions of SolO \citep{solarmach}, which was at a solar distance of 0.66 au, and other spacecraft in the solar ecliptic plane during the event. The coloured Parker spiral lines approximating the direction of the interplanetary magnetic field (IMF) from the Sun to different observers are plotted based on the measured solar wind speed (which was about 430 km/s at SolO around the SEP onset time; 680 and 700 km/s at Earth and STEREO A, respectively). 
The arrow shows the longitude of the C2.6 flare which approximates the associated CME eruption and its propagation direction. The dashed line represents the best magnetic connectivity to the flare based on Earth-observed solar wind speed.

Figure~\ref{fig:1109}(c) shows combined in-situ measurements at SolO including energetic particles, solar wind plasma and IMF. The first panel presents the intensity-time profile for protons at different energy bins.
The second panel shows the dynamic spectrum of proton flux covering a continuous energy range from $\sim$50 keV up to 105 MeV. 
The SEP event starts at 13:32 UT on 2023-11-09,  which is the onset time of the first arrival energy, indicated by the red arrow at around 3 MeV.
This plot shows a clear IVD component at energies above a few MeV in addition to the normal VD part at lower energies. More detailed analysis of SEP transport based on the two components is shown in Fig.~\ref{fig:1109ds} and discussed later. 

Figure~\ref{fig:1109}(c) shows that at the onset of the SEP event, most plasma and magnetic conditions were in a relatively quiet state. About two days later, the in-situ plasma information provides the signature of a shock arrival, as indicated by the red solid line. Considering the timing and speed of the shock, as also supported by the EUHFORIA simulation (available in SI), this should have been the shock associated with the SEP event. Upon the shock arrival, energetic storm particles (ESPs), which are trapped near the shock by self-generated waves \citep{reames1999}, were also seen in energy channels below a few MeV. 
In the sheath and within the CME, fine structures of the SEP flux are nicely seen which may be related to rapid changes of magnetic structures in the solar wind \citep{wimmer2023a}. 

\subsection{SEP release time and path length}\label{sec:1109ds}

\begin{figure*}
    \resizebox{\textwidth}{!}
    {\includegraphics[height=0.05\textheight]{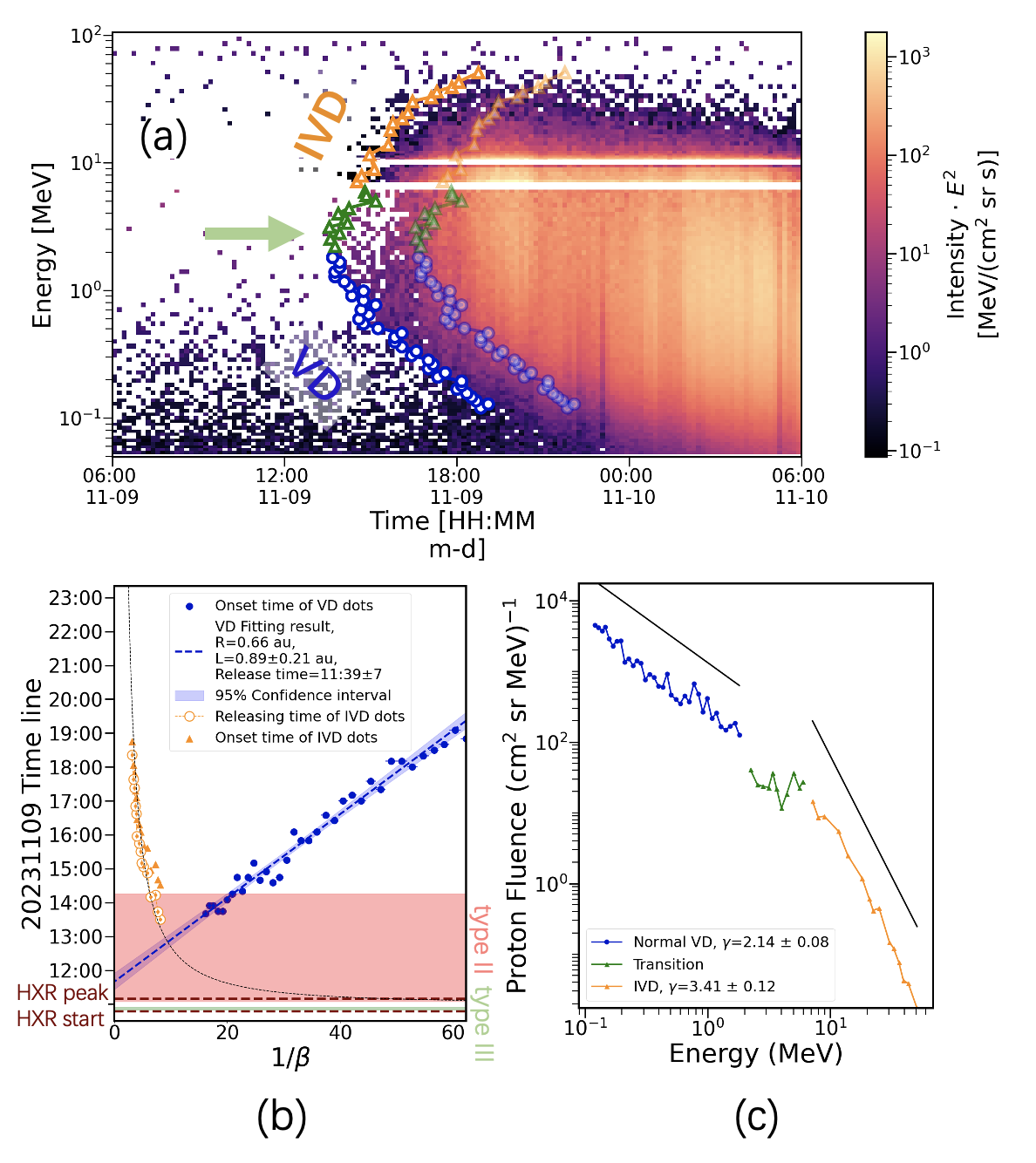} }
    \caption{(a) Dynamic spectra of the early phase of the 2023-11-09 event; (b) The release time and path length derived from VDA of protons below $\sim$2 MeV and from IVD release time analysis of protons above $\sim$7 MeV and (c) The 3-hour-integrated proton energy spectra starting from the onset of each energy range (between the two set of markers in (a)).
    In (a), the onset time for VD particles is marked with blue circles, while the onset time for IVD particles is indicated with orange triangles (see Section Methods for onset time determination. The big green arrow indicates the first arrival energy within the transition range between VD and IVD energy, which is marked with green triangles.
    The onset time is used in panel (b) to derive the release time and path length of energetic protons shown in the legend (Eqn.~\ref{eq:VDAclassic} and \ref{equ:ivd_release} in Methods). The theory-fitted release time based on DSA theory is plotted as the dotted line.
   Timing of the C2.6 flare hard X-ray ($\sim$10 keV) is marked by horizontal dashed lines; Type II and Type III radio bursts shifted to the solar surface (subtracting 8.33 min accounting for the time that photons need to reach Earth) are shown in pink and green bands, respectively. In (c), the power-law fitting obtained for the low- and high-energy parts is marked by the black lines and the legends. 
     }
    \label{fig:1109ds}
\end{figure*}

Figure~\ref{fig:1109ds} illustrates the typical analysis performed to obtain the release time and path length derived from the velocity dispersion analysis \citep[VDA,][]{Prise2014} of protons under $\sim$2 MeV as well as the release time of IVD particles with different energy above $\sim$ 7 MeV. VD and IVD analyses follow the methods described in Section Method. 
The dynamic spectrum in panel (a) shows a triangular structure with low-energy VD component and high-energy IVD part for the 2023-11-09 event. The transition energy is around 3 MeV (marked by the green triangles between 2 and 7 MeV), i.e., these are the particles arriving earliest at SolO and have very low counting statistics. Besides, it is not clear which component (VD or IVD) they belong to. For this reason, we have ruled out particles in this energy range in the following study. 

The VDA results are shown in Figure~\ref{fig:1109ds}(b) by the blue points and the linear fit with the release time $t_{\text{0}}$ derived as 11:39$\pm$7 UT which is broadly consistent with the 
hard X-ray (HXR) peak time and the start of type II radio emission (both were 11:05 UT).
This indicates that these particles are possibly accelerated during the flare reconnection process and/or the initial phase of the shock when it was still in the low corona.
Notably, the VDA-derived path length is longer than the corresponding Parker spiral length (also for the other two events shown in SI and Table \ref{tab:results}). This supports that energetic particles may experience transport effects such as scattering or cross-field propagation\citep{jokipii1966,reames1999} and also that the Parker spiral approximation is an over-simplified situation which has been challenged by recent studies \citep{Laitinen2015,kasper2019}.

The IVD structure indicates that these particles were not released at the same time and that particles with higher energies were released later, a scenario more likely related to shock acceleration rather than flare acceleration processes (with further explanations provided in the Discussion). We derive their energy-dependent release time starting from an initial guess that they followed the same path length as the VD particles. This path length is further iterated taking into account the shock propagating away from the Sun thus reducing the particle paths to the observer (see more details in Method and Equation \ref{equ:ivd_release}). The derived release time $t_{\text{release}}(E')$ for each energy $E'$ is marked in Fig.~\ref{fig:1109ds}(b) by orange circles which clearly show that higher-energy protons (to the left of x-axis) are released later than lower-energy ones. 

If VD and IVD particles do come from different sources and different acceleration processes, their energy spectra may be different.
In order to resolve the energy spectrum most similar to that upon the release, we use particle flux during the initial three hours following the onset of each energy range featuring particles that experienced least scattering \citep{Saiz2005}. We then deduce the energy spectra of VD and IVD particles, respectively, and find that they show different power-law indices (Figure~\ref{fig:1109ds}c).
For reference, two black lines are shown with power-law indices of 2.14 and 3.41 for low-energy VD and high-energy IVD parts, respectively. 

SEP spectra often show double power-lawfeatures, but their origin is still under debate, possibly due to transport effect \citep{li2015scatter, Wang2024} or as a direct consequence of the time-dependent shock acceleration process, which can reflect properties of the shock geometry \citep{mason2012,li2009b,desai2016}. 
We suggest that, in this event, the double power-law feature indicates that there may be two different acceleration phases, e.g., DSA for the high-energy part as discussed in detail later while flare reconnection or early-stage-shock acceleration for the low-energy part, consistent with the results indicated by the release time analysis in Figure~\ref{fig:1109ds}b. 

\section{Discussion}

Interplanetary shocks are the main source of high-energy protons in the inner heliosphere causing the commonly-known gradual SEP events \citep{Reames2013}. 
Assuming that the IVD particles are mainly accelerated at the shock while it propagates outwards, we interpret the origin of IVD feature exploring two different scenarios (which may also be co-responsibly contributing agents). 

\subsection{Acceleration time evolution during diffusive shock acceleration}\label{sec:theory_dsa}

Among various particle acceleration mechanisms in the heliosphere, diffusive shock acceleration (DSA) is one of the most common and important mechanisms \citep{vainio1999,Drury1983}, especially for parallel and quasi-parallel shocks. 
DSA can be explained through the Fermi acceleration process, during which particles could cross the shock multiple times by elastically scattering off magnetic irregularities that converge at the shock. Particles gain a small amount of energy in each traverse of the shock front. Strong turbulent magnetic fields help the shock to trap particles, which can be continuously accelerated, while the final energy of the escaping particle is determined by its initial energy and the duration time of the acceleration as well as the shock properties \citep{vainio1999}.
Bell \citep{bell1978} and Drury \citep{Drury1983} developed individual particle approach to compute the energy (or momentum) gain of a particle from one crossing through the shock, and Jones \citep{jones1991} proposed a simpler way to obtain this value, just combining the equation for the conservation of particles:

\begin{gather}
\frac{\partial}{\partial x}(j_x)+\frac{\partial}{\partial p}(j_p)=0,
\end{gather}

where $j_x$ and $j_p$ are the fluxes of particles along the x-(shock normal direction) and p-axes (particle momentum) respectively, together with the diffusion-convection equation below \citep{ellison1990} 

\begin{gather}
\frac{\partial}{\partial x}\left[uf(x,p)-\kappa\frac{\partial f(x,p)}{\partial x}\right]=\frac{1}{3}\left(\frac{\partial u}{\partial x}\right)\frac{\partial}{\partial p}[pf(x,p)],
\end{gather}

which leads to the final average momentum of a particle with initial momentum $p_0$ after crossing the shock $N$ times as:

\begin{gather}\label{eq:multi}
<p>(N)=\prod_{i=1}^N[1+\frac{2}{3}(u_u-u_d)/v_i]p_0,
\end{gather}

where $u_u$ (or $u_d$) is the upstream (or downstream) flow speed in the shock frame, and $v_i$ is the particle speed. 

Based on magnetic field data across the shock front when it arrived at SolO (see the solid vertical line in Fig.~\ref{fig:1109}c) of 2023-11-09 event, we derive that the angle between the shock normal and upstream magnetic field $\theta_{Bn}$ is about $20^{\circ}$, i.e. the shock qualifies as a quasi-parallel shock, indicating that DSA may likely have been the main acceleration mechanism during this long-lasting SEP event, if the shock direction did not evolve significantly over the distance. 

There is no easy way to directly resolve the shock properties for accelerating first-arriving IVD particles when the shock was closer to the Sun. We here use the accelerated IVD particle spectral index (Fig.~\ref{fig:1109ds}c) to derive the downstream speed when it was accelerating the IVD particles based on the method derived by Bell \citep{bell1978} and Jones \citep{jones1991}. In detail, they considered the probability of particles crossing the shock $N$ times and used this to derive the particle distribution and spectrum. Assuming an isotropic particle distribution in the local plasma flow frame, one can calculate the probability that a particle, having once crossed the shock in the configuration space, will return. Additionally, the probability that a particle returns at least N/2 times can be determined, which is equivalent to the probability of the particle crossing the shock at least $N$ times.

The density function can be given by the partial differentiation of probability function to the momentum:

\begin{gather}\label{eq:f}
f(p)=\left(\frac{n_0}{p_0}\right)\left(\frac{3r}{r-1}\right)\left(\frac{p}{p_0}\right)^{-\sigma}
\end{gather}

where $p_0$ and $p$ are the initial and final momentum of the particle (same as defined before), $r$ is the compression ratio of the shock: $r\equiv \rho_d/\rho_u=u_{u}/u_{d}$ and $\sigma\equiv (r+2)/(r-1)$ , and $n_0$ is the upstream number density. Equation~\ref{eq:f} presents the well-known form of the accelerated particle energy spectrum, exhibiting a power-law behaviour of $dJ/dE \propto E^{-\sigma/2}=E^{-\Gamma}$. Here, the compression ratio $r$ is constrained to $1<r<(\gamma_G+1)/(\gamma_G-1)$ for a non-relativistic monoatomic gas, where $\gamma_G=5/3$. Consequently, the upper limit of $r$ is 4 so that the energy spectral index $-\Gamma$ should be larger than 1. 

Based on the derived particle spectral index $-\Gamma$ and the measured upstream flow speed $u_u$ (in the shock frame), we can determine the compression ratio $r$, and subsequently, downstream flow speed can be given as 

\begin{gather}\label{ud}
u_d=\frac{(2\Gamma-1)u_u}{2+2\Gamma}. 
\end{gather}


Applying the method to the 2023-11-09 event, the in-situ measured upstream flow speed is $u_u=-542$km/s (perpendicular to the shock surface and in the shock frame) upon the shock arrival. The spectral index of the IVD onset particles is 3.41 which corresponds to a shock compression ratio of $r=1.5$ in DSA theory. We then derive the downstream flow speed $u_d=-358$ km/s when the shock was close to the Sun. 
In the above calculation, we have assumed the shock properties and the seed particle spectra do not change over the distance from about 0.05 to 0.14 au. Future investigations can consider the evolution of shock properties and the seed particles so that IVD spectrum actually combines injection sources at different times. 

Two additional events on 2023-12-24 and 2023-12-31 also have clear IVD features with sufficient remote-sensing data support. We also studied them with the results shown in Table~\ref{tab:results}. 

\begin{figure*}
    \centering
    \resizebox{\textwidth}{!}
    {\includegraphics{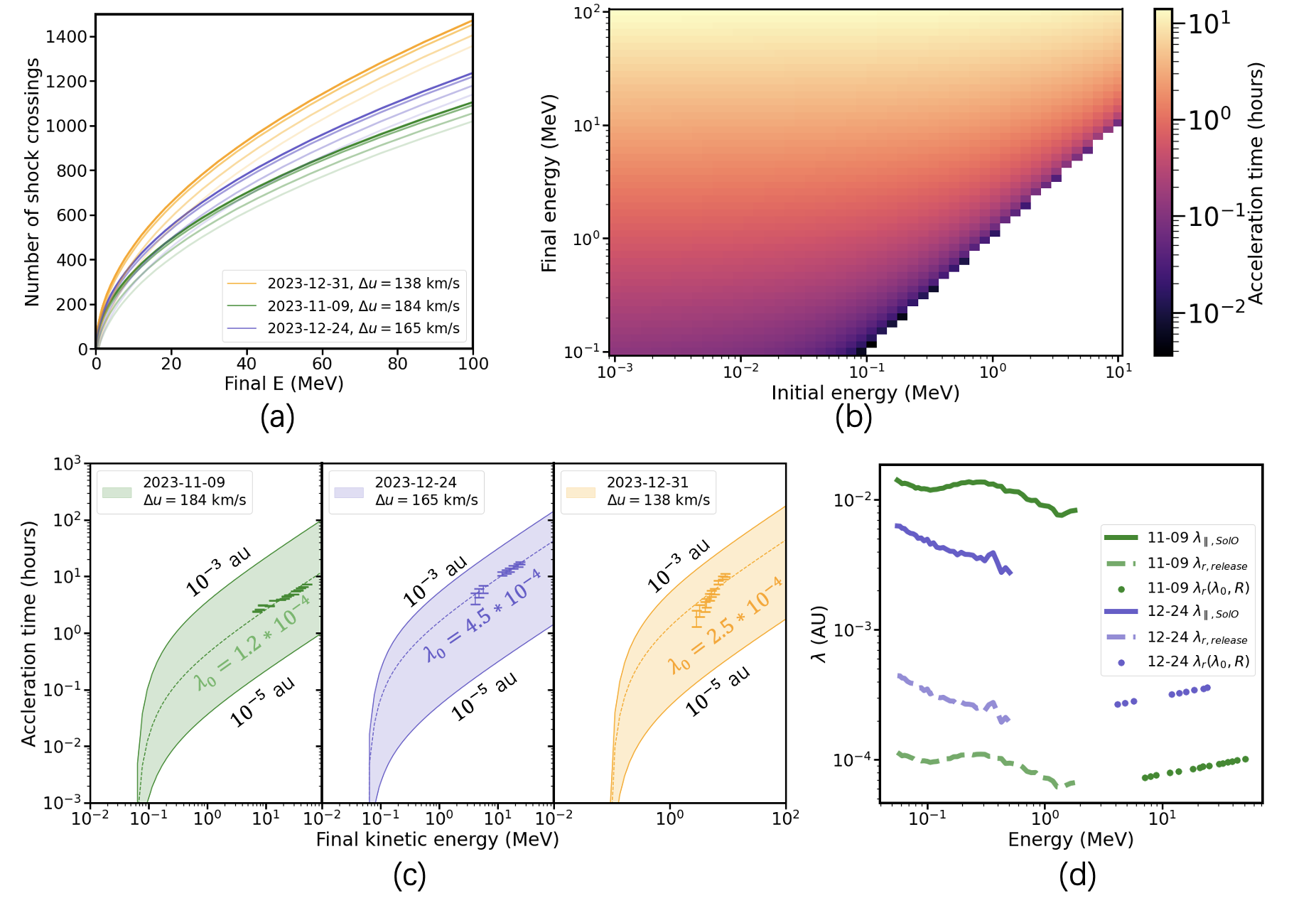}}
    \caption{(a) Shock crossing times versus final proton energy derived from different events with different shock properties (shown as different $\Delta u$ in the legend). For each event, four lines transitioning from darker to lighter represent an increase of initial kinetic energy which is 10$^{-3}$, 10$^{-2}$, 10$^{-1}$ and 1 MeV, respectively. 
    (b) Acceleration time matrix for the 2023-11-09 event. Colour scale represents the acceleration time $\tau_a$ required to accelerate a particle from an initial energy (x-axis) to a final energy (y-axis), with Eqn.~\ref{eq:acc time} using $\lambda_0=1.2 \cdot10^{-4}$ au. The matrices for other events and other $\lambda_0$ values have similar structures and are not shown here. 
    (c) $\tau_a$ versus the final proton energy for three different events. Each coloured band corresponds to the possible solutions of $\tau_a$ with $\lambda_0$ ranging from $10^{-5}$ to $10^{-3}$ au. The markers with error bars are results derived from the observations based on IVD analysis. 
    (d) The upper two lines are the parallel mean free path $\lambda_\parallel$ derived from the ESP flux when the shock arrived at SolO for the 2023-11-09 and 2023-12-24 events (while the 2023-12-31 event had a contamination by a following SEP event). The lower dashed lines are $\lambda_{r, release}$ at the acceleration source region derived from the upper ones considering the radial gradient of $\lambda_\parallel$. The value of $\lambda_\parallel$ obtained from fitting the observations in panel (c) is plotted (above $\sim$ 7 MeV as dots) as a reference. See the text for more explanations. }
    \label{fig:acc time}
\end{figure*}

The number of times a particle needs to traverse the shock front, $N$, to accelerate from its initial momentum $p_0$ to a specified momentum $p$ is obtained via Equation~\ref{eq:multi} shown above. 
It also shows that the difference of the upstream and downstream shock speed, defined as $\triangle u = \lvert u_u-u_d\rvert$  can influence the energy gain: larger $\triangle u$ would result in more efficient acceleration. 
Fig.~\ref{fig:acc time}(a) shows the required number of crossings for protons reaching different energies with the $\triangle u$ derived from the 2023-11-09 event and another two selected events. 
For each event, lines that transition from darker to lighter colours represent an increase in the initial $p_0$.

A more quantified acceleration time scale can be obtained by transiting from a single-particle approach to a macroscopic perspective.
Drury \citep{Drury1983} first derived the mean acceleration time from the steady, time-dependent solution of the transport equation, under the condition that a mono-energetic source of particles $Q\delta (p-p_0)$ is available at $t=0$, and the distribution of accelerated particles is $f(t,x,p)=0$ at $t=0$. The expression for the mean acceleration time is derived as:

\begin{gather}\label{eq:acc time}
\tau_a = \frac{3}{u_u - u_d} \int_{p_{\text{inj}}}^{p} \kappa_{rr} \left( \frac{1}{u_u} + \frac{1}{u_d} \right) \frac{dp'}{p'}
\end{gather}

where $p_{\text{inj}}$ is the initial momentum of the source particles, and $\kappa_{rr}$ is the effective diffusion coefficient along the shock normal in both upstream and downstream regions \citep{prinsloo2019} related to radial mean free path $\lambda_{rr}$ by $\kappa_{rr}=\frac{v}{3} \cdot \lambda_{rr}=\frac{v}{3} \cdot \lambda_0 \left (\frac{R}{R_0} \right )^{1/3}$
where $v$ is the particle speed, $R$ is the particle rigidity and $\lambda_0$ is a reference mean free path at $R_0\equiv 1$ GV. $\lambda_0$ indicates the ability of the shock to confine particles: smaller values of $\lambda_0$ lead to better confinement and more efficient acceleration of particles. The rigidity dependence ($R^{1/3}$) follows the derivation by Jokipii et al. \citep{jokipii1966}. However, we acknowledge that the spectral index of amplified turbulence near shocks can deviate from the Kolmogorov value \citep{li2005}, resulting in different rigidity dependencies of the diffusion coefficient. In reality, $\lambda_{rr}$ should be different across the shock \citep{Drury1983,Jokipii1986}, the effective $\kappa_{rr}$ here represents the value in the upstream region where the scattering time is much longer. 

Figure~\ref{fig:acc time}(b) shows the acceleration time matrix for the 2023-11-09 event based on the above theory (using the best fit $\lambda_0$ derived in panel c). 
The result is consistent with that shown in Figure~\ref{fig:acc time}(a).
As indicated by the colour gradient, the required acceleration time does not vary significantly with initial energy, but increases substantially with final energy. That is, protons with higher energies need more time to be accelerated.

Figure~\ref{fig:acc time}(c) shows the acceleration time taking into account the actual event spectra. Specifically, we consider the seed particles without a single energy, but rather with a distribution similar to the VD particle spectrum (blue dots in Fig.~\ref{fig:1109ds}c) and are accelerated to have a final spectrum of the IVD particles (orange dots). Here we have assumed that the spectra of first-arriving particles may represent the spectra close to the acceleration site as they experienced least scattering. In fact, the seed particle energy is not important for the results while the final energy is more critical (Figure~\ref{fig:acc time}b). Moreover, transport effects such as adiabatic cooling could modify the particle spectra. But the effect is only significant for low-energy particles below about 10 MeV \citep{Wang2024} while the IVD particle spectrum is mostly above this energy range. 
Meanwhile, the acceleration time also depends on $\lambda_0$ as shown by the range of the coloured band. 
Generally, a smaller mean free path results in a shorter acceleration time for a given final energy, implying that shocks that better confine particles can better accelerate them.

Type II radio bursts are signals of shock particle acceleration; therefore, the possible initial time of DSA acceleration ($\tau_0$) can be approximated as the start of the type II radio emission \citep{Reames2021}.
However, for the events on 2023-12-24 and 2023-12-31, the only possible CME associated with the SEP event occurred shortly before the flare which was closest to the $t_0$ derived from VDA. A possible scenario to explain this ``confusing" situation is that the flare-accelerated particles served as seed particles that were further accelerated by the shock which erupted earlier \citep[e.g.,][]{lizank2005}. Thus, it is also plausible to assume the flare peak time as $\tau_0$. 
We therefore consider both possibilities (type II start, flare peak) as $\tau_0$ and deduce the required acceleration time of the SolO-observed IVD particles as $\tau_a(E') = t_{\text{release}}(E^{'})-\tau_0$ where $t_{\text{release}}(E^{'})$ is the derived release time (Fig.~\ref{fig:1109ds} (b)).
Results of $\tau_a(E')$ derived from two possible choices of  $\tau_0$ are plotted as error bars in Fig.~\ref{fig:acc time} (c) and the data can be fitted with a fixed $\lambda_0$ (listed in Table~\ref{tab:results}) shown by the dashed lines.
The theoretical release time ($\tau_0$ plus $\tau_a$ fitted from $\lambda_0$) for the first event is also plotted in Figure \ref{fig:1109ds}(b) as the black-dotted curve which agrees nicely with the observation-derived release time (orange circles).

All three events show consistent values of $\lambda_{0}$ around 10$^{-4}$ au. Although more statistics are needed for a more general conclusion, this indicates common characteristics of the DSA process, e.g., $\lambda_0$ is determined by the excitation of Alfvén waves near the shock front by streaming protons\citep{li2005}. Consequently, $\lambda_0$ near the shock front derived here is much smaller than that under quiet solar wind conditions \citep[0.4 AU derived by Li et al.][]{li2005}. We further derive diffusion coefficient $\kappa$ from $\lambda_0$ to be $\sim 10^{13}$ $m^2/s$ between 1 MeV and 100 MeV. The values are generally a few times higher than those derived by Li et al. at 0.18 au for 1-10 MeV protons under shock conditions, while the results are more comparable at 100 MeV. Further investigations are needed for understanding the difference of $\kappa$ derived from two independent methods.

For the 2023-11-09 and 2023-12-24 events, we can compare the derived $\lambda_{0, release}$ ($< \sim 0.3$ au, see Table \ref{tab:results}) with the in-situ mean free path obtained at the shock when it arrived at SolO. 
Following the e-folding time method derived from ESP flux \citep{beeck1989}, we obtain the local parallel mean free path $\lambda_{\parallel, SolO}$ in the upstream shock region at SolO. 
The result is shown as solid lines in Fig.~\ref{fig:acc time}d where $\lambda_{\parallel, SolO}$ ranges in 0.003-0.02 au for two events and varies slightly with energy. 
The radial mean free path can be written as $\lambda_r=\lambda_\parallel \cos ^2 \Psi$ where $\Psi$ is the angle between the local magnetic field direction and the radial direction. Based on the IMF observations, we calculate $\cos ^2 \Psi$ as approximately 0.5. 
Wang et al. \citep{Wang2022} considered the radial dependence of the particle's diffusion coefficient as $\kappa \propto D^\beta$ with $D$ being the distance to the Sun and $0\leq\beta<2$. Consequently, $\lambda_{r, release}$ near the Sun (at the release site) can be deduced from the local $\lambda_{\parallel, SolO}$ at the location of SolO shown as dashed lines in Fig.~\ref{fig:acc time}d.
Compared to $\lambda_r$ derived from the acceleration time (dots), $\lambda_{r, release}$ can be aligned to the same magnitude (although at lower energies) with $\beta\sim$1.6 and $\beta\sim$0.8 for 2023-11-09 and 2023-12-24 events, respectively (see Table~\ref{tab:results}). 
These results are roughly in agreement with Chen et al. \citep{chen2024} who obtained the radial gradient of the parallel diffusion coefficient $\kappa_\parallel \propto D^{1.17}$ using the measured magnetic turbulence power spectra in the inner heliosphere. 
The difference between our results and theirs can be attributed to the difference between quiet IMF condition and the turbulent conditions at the shock front. 
Note that calculation of in-situ $\lambda_\parallel $ using ESP observations for 2023-12-31 event is not possible because another SEP event has occurred shortly after the onset of this event.

It is worth mentioning that DSA mechanism could also be applied to gradual plasma compressions without a true shock discontinuity \citep{giacalone2002, giacalone2005}. Observational evidence has revealed energetic particle events potentially associated with such compressed regions, even in the absence of strong CMEs or shocks \citep{malandraki2023}. If the development of shock formation is slow, compression wave acceleration can also contribute to the observed IVD feature in these events. 

\subsection{Connectivity change during the shock propagation}

\begin{figure*}
    \centering
    \resizebox{\textwidth}{!}
    {\includegraphics{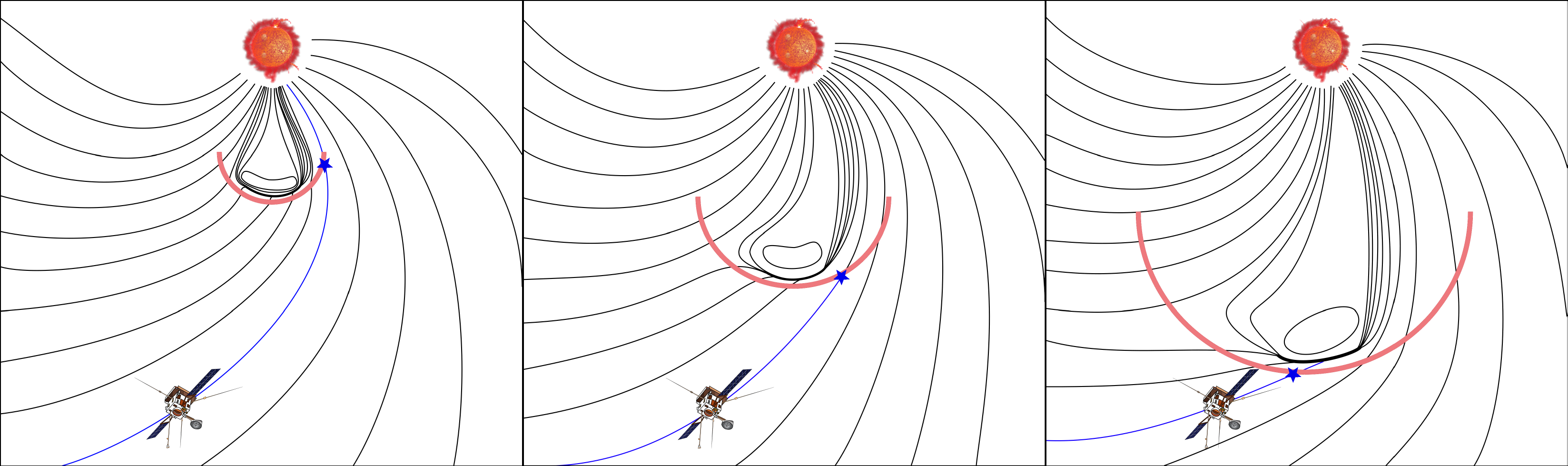}}
    \caption{Sketch depicting the connectivity scenario which could potentially explain the later incremental arrival of higher energy protons to the observer. The blue star marks the cobpoint, which is a point at the shock (marked by the red curves) that is magnetically connected to the observer. See the text for more explanations.}
    \label{fig:connection}
\end{figure*}

We also investigate the alternative explanation on how the evolving magnetic connection between the observer and shock front may cause the observed IVD.
Under nominal solar wind conditions, the point at the shock that is magnetically connected to the observer, defined as the cobpoint \citep{Heras1995}, moves along the front of the shock eastward during its propagation. During this process, the observer is connected to different regions of the shock which may have different acceleration efficiencies \citep{Pomoell2015}. 
Figure~\ref{fig:connection} shows a sketch of the simplified scenario for the 2023 November 9 event, depicting the evolution of the shock (red curve) driven by the CME (simplistically represented as bubble). The cobpoint (blue star) slides along the shock front over time with changing magnetic connection to the observer (SolO), depicted by a blue curve in each panel.

The acceleration efficiency of the shock is generally believed to be higher close to the shock nose while it decreases towards the shock flank, although detailed modelling of the shock evolution shows that the shock parameters (Mach number, compression ratio, etc.) may change significantly both spatially and temporally as the CME propagates through the heliosphere \citep{Jin2022}. 
Assuming the simple case that the early-connected shock flank is less efficient in SEP acceleration while the later-connected shock nose is a more efficient accelerator, Figure~\ref{fig:connection} provides a possible explanation for the observed IVD feature in the 2023-11-09 event where the CME width and direction are derived based on GCS fitting.

Considering the connectivity changing from the flank to the nose, we check the relative angular information of all events. Table~\ref{tab:results2} shows the separation angle between SolO and the solar source (i.e., the flare associated with the SEP) and the calculated angle between SolO's magnetic footpoint and the source. For both angles we observe that IVD features are not restricted to a specific range of longitudinal separations, thus implying that the connectivity model may not serve as a general explanation for all events.
Besides, for events with a short duration of IVD and a small radial distance ($< 0.2$ au) upon the derived release time, the role of connectivity change should not be significant. 

Furthermore, to quantify the observed IVD features, we would need a combined approach to model the shock evolution in both time and space as well as particle acceleration and transport processes \citep{Li2021}. 
Kouloumvakos et al. \citep{Kouloumvakos2025} have made such an attempt to model the event PSP observed on 2022-09-05 at about 15 solar radii which was reported first by Cohen et al. \citep{Cohen2024} who suggested that the later released higher-energy particles do not have enough time to overtake the earlier released lower-energy ions before reaching the spacecraft which was sufficiently close to the Sun. Kouloumvakos et al. modelled both the shock and particles simultaneously and further proposed that PSP was initially connected to the weaker part of the shock and later to a strengthened shock with higher acceleration efficiency. The connectivity change has resulted in the observed later arrival of particles with higher energies at PSP. But the modeled particle intensity was higher than observed and particle profile detected at Solar Orbiter was not well reproduced. More consistent Sun-heliosphere-shock-particle modelling effort would be necessary to better reveal the general nature of IVD particles, such as that by Ding et al. \citep{ding2025} who successfully reproduced the observed IVD signatures by SolO on 2022-06-07.

Additionally, as we noted for the 2023-11-09 event, two CMEs were launched within a short time window. Interacting CMEs may result in possible changes in the shock angle \citep{lugaz2017} and thus impact the evolution of the connectivity, which becomes a much more complex scenario. In fact, we cannot rule out the presence of a shock driven by the northern CME, which may have interacted with the main southern shock, giving rise to a more complex particle acceleration process due to shock interactions \citep{zhao2014}.
Such a modelling effort would require many input parameters that are currently not easy to validate against observations and is thus not pursued further in this study.

\subsection{Conclusion}
In summary, this study shows a new type of solar energetic proton events observed by SolO which has an unusual inverse velocity dispersion structure (at energies above a few MeV) during the event onset in addition to the typical velocity dispersion at lower energies. 
This phenomenon can be explained to be caused by the delayed release of particles with increasing energies. 
We analysed the travel path and release time of particles for both VD and IVD components and have quantified the IVD release time as a function of proton energies. 
We found that the VD proton release time is consistent with the flare X-ray bursts and initial Type II radio emission, which indicates that these particles are likely associated with the flare process or early-shock formation close to the Sun.
For IVD protons, the release time is much later when the shock is at a solar distance between about 0.05 and 0.2 au and it increases with particle energy. 

These observations can be explained by two different mechanisms (or a combination of them). 
First, it takes longer time for higher-energy particles to be accelerated as required by diffusive shock acceleration mechanism, based on which we derived the shock parameters during the acceleration process.
Second, as the shock propagates outward, the observer's magnetic connection to the shock front changes. In the case of an IVD event, the observer gets connected to regions with increasing acceleration efficiencies. However, for the events studied here, the connectivity theory may not be a general explanation as it requires restricted separation angles between the observer and the source. 

To conclude, this new trait in the high-energy range of solar proton events helps us to innovatively derive physical parameters during the acceleration process directly from observations, in particular the actual acceleration time scales that cannot be observed directly.
Nevertheless, more coordinated observation and modelling efforts are needed to constrain the two different mechanisms, their interlink and contributions to the observations. In particular, shock-related acceleration can be modeled with MHD shock modelling coupled with particle transport simulations \citep{ding2025}. Such models can provide key parameters including evolution of shock properties, magnetic connectivity, and energy-dependent release profiles, which can then be directly compared to the in-situ particle and plasma observations.
We emphasize that particle radiation environment during solar eruptions is dynamically varying in both spatial and temporal dimensions and that radiation assessment should be tailored for different observers taking into account the relative position of the observer to the acceleration source and the evolution of the source. 
Last but not least, IVD features have not been reported before recent observations because they may have been overlooked, so the development of the new instrumentation with better time and energy resolution has opened new opportunities to discover new phenomena in the observations.

\section{Methods} \label{sec:method}
\subsection{Velocity Dispersion Analysis}\label{sec:VDA}
Under the assumption that at the beginning of the event, the first-arriving particles are released simultaneously and travel along the same path length, we expect that particles with higher energies would arrive earlier. This has been frequently observed before and is called the velocity dispersion \citep{Krucker99, Prise2014, xu2020first, wimmer2023a}. 
The VDA often assumes a small cross-field scattering process which is supported by the high anisotropy of the SEPs upon the onset of this event, as shown in the supplementary information.
To determine the release time and path length of particles with the velocity dispersion feature, the following function is often applied:

\begin{gather}
   t_{\text{onset}}(E)=t_{\text{0}}+8.33\frac{\text{min}}{\text{au}}\cdot \frac{L_0}{\beta(E)},
   \label{eq:VDAclassic}
\end{gather}

where $t_{\text{onset}}(E)$ (in the unit of minute) is the observed SEP onset time (as explained later) for particles with kinetic energy E, $\beta (E)=v(E)/c$ is related to the velocity of the particles, $t_{\text{0}}$ (in the unit of minute) and $L_0$ (in the unit of 1 au) are the initial release time and the travel path to be derived from this function. 
Based on $\beta (E)$ (x-axis) and observed $t_{\text{onset}}$ (y-axis), we can linearly fit the above function, using orthogonal distance regression method. The uncertainty in energy arises from the energy bin width of EPD, while the uncertainty in the time originates from 5-minute intervals of particle flux data. This method allows us to determine the $L_0$ as the slope and $t_0$ as the intercept with the y-axis.
In the above process, the Cumulative sum (CUSUM) quality-control schemes \citep{Heikinmaa2005} was applied to determine the onset time $t_{\text{onset}}$ for particle fluxes at each energy of the VD part.

Based on determined $t_0$, we identify the flare and CME which occurred nearest (normally within one hour) as the most-likely responsible cause of the SEP event (also see SI for the flare and CME observations). 

\subsection{Inverse Velocity Dispersion Analysis}\label{sec:IVD}
For particles with the inverse velocity dispersion (IVD) features, we assume that their release time $t_{\text{release}}(E^{'})$ and path length $L(E^{'})$ change as a function of energy (expressed as $E^{'}$ to be different from $E$ used for VDA). 
The particle release time $t_{\text{release}}(E^{'})$, onset time $t_{\text{onset}}(E^{'})$, velocity $\beta(E^{'})$ and propagation distance $L(E^{'})$ still follow the expression below:

\begin{gather}
   t_{\text{onset}}(E^{'})=t_{\text{release}}(E^{'})+8.33\frac{\text{min}}{\text{au}}\cdot \frac{L(E^{'})}{\beta(E^{'})},
   \label{equ:ivd_release}
\end{gather}

However, for IVD particles, the low signal-to-noise ratio and inadequate data during the pre-event background period often result in poor performance and unreliable results of the Poisson-CUSUM algorithm which was used for VDA. 
Here we opt to manually determine the onset time based on the 2-d histogram of the energy-dependent particle flux. To minimize uncertainties, this process was repeated multiple times until the result is stabilised.

With both $t_{\text{release}}(E^{'})$ and $L(E^{'})$ being variables, it is not feasible to directly fit the above function. So we use an iteration process to solve the above function. We assume the initial value of $L(E^{'})$ equals to $L_0$ which is derived from Equation~\ref{eq:VDAclassic} and calculate $L(E^{'})$ as the CME shock propagates away from the Sun. 

First, based on the initial properties of the CME derived from remote-sensing observations, we use the Drag Based Model \citep[DBM,][]{vrsnak2013} to propagate the shock from 17 solar radii into the interplanetary space. 
The DBM is a semi-empirical tool for CME propagation, assuming that beyond $\sim$15 solar radii, the dynamics are governed solely by the interaction between the CME and the ambient solar wind. Despite of its simplicity, the model has been proven to perform equally well compared to other more complex MHD models \citep{vrsnak2014, Dumbovic2021}. 
Detailed input CME and solar wind parameters for the model are given in SI.

Second, with the consideration that the particle release site (i.e., the shock front) evolves with time (as does energy $E^{'}$) when the shock moves outward, we then derive the shock distance at different release time $t_{release}(E^{'})$ with the path length $L(E^{'})$ modified at each $t_{release}(E^{'})$ using an iteration process as detailed below. 

\begin{enumerate}
    \item In the first step, we assume that $L(E^{'})$ is approximately $L_0$ as derived from Eqn.~\ref{eq:VDAclassic}. Then Eqn.~\ref{equ:ivd_release} can be solved to obtain $t_{release}(E^{'})$ for $E^{'}$. 
    \item At the above derived time $t_{release}(E^{'})$, we obtain the shock propagation distance from the Sun $R(t)$ using DBM. We then modify the particle propagation path as $L(E^{'})-R(t_{release}(E^{'}))$ to account for the shortened distance of the particle propagation path compared to that of the initial release (see further explanations later). 
    \item With the modified path, we re-calculate $t_{release}(E^{'})$ from Eqn.~\ref{equ:ivd_release} which corresponds to another $R(t)$ that can be different from the previous step (item 2) that will further change the propagation path. 
    \item The above steps (items 2-3) are repeated until $t_{release}(E^{'})$ and $L(E^{'})$ converge (i.e, newly derived values equal to those derived in the previous step).
    \item The uncertainty of $t_{\text{release}}(E^{'})$ is calculated using error propagation method accounting for the observational time interval and energy range.
\end{enumerate}
In the above approach, we assumed that the propagation distance of the shock corresponds to the shortened distance of the SEP path length due to this reason: for the derived release time $t_{release}(E^{'})$, normally the CME was still close to the Sun (mostly within 0.3 au) where the interplanetary magnetic field has a much higher radial component.

\begin{table*}[]
\centering
\resizebox{\textwidth}{!}{%
\begin{threeparttable}
\begin{tabular}{clccc}
\toprule
\textbf{Event dates}                       &                                      & \textbf{2023-11-09} & \textbf{2023-12-24} & \textbf{2023-12-31}  \\ \midrule
\multirow{11}{*}{\textbf{Observation}}     & Distance of SolO                     & 0.66 au             & 0.94 au             & 0.95 au              \\
                                           & Flare location                       & W04, S10            & W29, S19            & W89,N09    \\
                                           & Flare Class                          & C2.6                & M2.6                & C2.8\\
                                           & $\Delta \phi$                      & 21°
                                               & 7.2°                 & -46.4°    \\                                    
                                           & Flare start and end time (UT)                    & 10:45\,--\,11:29    & 16:29\,--\,16:48    & 11:47\,--\,12:17     \\
                                           & Flare peak time (UT)                   &  11:05               & 16:41               & 11:59                \\
                                           & CME eruption time\tnote{I} (UT)                    & 11:40               & 15:40               & 9:52                \\
                                           & CME starting speed                   & 870$\pm$87 km/s               & 740$\pm$74 km/s               & 830$\pm$83 km/s                \\                                    
                                           & Time of Type II radio burst (UT)        & 11:05\,--\,14:16    & 14:51\,--\,17:28    & 10:47\,--\,12:04     \\
                                           & $E_{0}$ (First arrival energy)                              & 3 MeV               & 1-4 MeV             & 2 MeV                \\
                                           & SEP onset                                & 13:32               & 23:09               & 15:15                \\
                                           & $E_{m}$ (Maximum energy observed by SolO)                              & 50 MeV              & 12 MeV              & - \tnote{II}                \\
                                           & $V_{SW}$ \tnote{III}                                & 350$\pm$70 km/s            & 300$\pm$60 km/s            & 400$\pm$80 km/s             \\
                                           & $\gamma_{_\text{VD}}$ (spectral index of the VD particles)                        & $-$2.14                & $-$1.94                & $-$0.61                 \\
                                           & $\gamma_{_\text{IVD}}$ (spectral index of the IVD particles)                       & $-$3.41                & $-$3                   & $-$3.66                 \\ \midrule
\multirow{5}{*}{\textbf{VD \& IVD Analysis }} & $t_{0}$ (UT) (derived from VDA)                   & 11:39 $\pm$ 7 mins  & 13:44 - 16:35 & 12:04 $\pm$ 6 mins   \\
                                           & Path length (derived from VDA)                         & $0.89 \pm 0.21$ au  & $1.61 \pm 0.45$ au  & $1.33 \pm 0.17$ au   \\
                                           & Parker spiral length                 & 0.72 au             & 1.04 au             & 1.09 au              \\
                                           & Releasing radial distance\tnote{IV}            & 0.05\,--\,0.16 au   & 0.14\,--\,0.38 au   & 0.07\,--\,0.2 au    \\
                                           & Slope of $E'$ vs. $t_{release}(E’ )$ & 9.25 MeV/h          & 2.88 MeV/h          & 0.69 MeV/h           \\ \midrule
\multirow{7}{*}{\textbf{Shock parameters}} & $u_{u}$ (shock upstream  speed)                             & $-$542 km/s           & $-$440 km/s           & $-$430 km/s            \\
                                           & Derived $u_{d}$ (shock downstream speed)                     & $-$358 km/s           & $-$275 km/s           & $-$291 km/s            \\
                                           & $r$ (shock compression ratio)                                 & 1.5                 & 1.6                 & 1.5                 \\
                                           & $\theta$ (shock normal angle)                            &  20°               &  25°               & -  \\
                                           & $\lambda_{0}$ at $t_{release}$ derived from IVDA    & $1.2\cdot10^{-4}$ au        & $4.5\cdot 10^{-4}$ au  & 
                                     $2.5 \cdot 10^{-4}$ au      \\
                                           & In-situ upstream $\lambda_{\parallel}$\tnote{V}              & $(0.8 -2)\cdot10^{-2}$ au   & $(2 - 6) \cdot10^{-3}$ au          & -  \\
                                           & The derived radial dependence of $\lambda_{\parallel}$ on solar distance D         & D$^{1.6}$                       & D$^{0.8}$           & -    \\ \bottomrule
\end{tabular}%

 \begin{tablenotes}
        \footnotesize
        \item[I] First appearance in C2. The time of the flares and CMEs is the observation time subtracting the time of light travelling from the Sun to SolO. 
        \item[II] The second SEP event caught up at SolO contaminating the current SEP IVD structure.
        \item[III] In-situ solar wind speed averaged over 10 hours before the shock arrival.  
        \item[IV] The modelled distance of the shock upon the derived IVD particles' release time.
        \item[V] The parallel mean free path derived from the e-folding time method.
\end{tablenotes}

\end{threeparttable}
}
\caption{Observations and analysis results of three different SEP events with clear IVDA features and their associated solar eruptions.}
\label{tab:results}
\end{table*}

\begin{table}[]
\centering
\resizebox{\textwidth}{!}{%
\begin{threeparttable}
\begin{tabular}{ccccccccc}
\hline
  \textbf{No.} &
  \textbf{Start Date} &
  \textbf{R$_{Sun-SolO}$ } &
  \textbf{Onset E } &
  \textbf{Max. E } &
  \textbf{Onset}  &
  \textbf{IVD Duration} &
  \textbf{$\delta_{SolO-source}$ } &
  \textbf{$\delta_{footpoint-source}$} \\ 
  & [yyyy-mm-dd] & [AU] & [MeV]   & [MeV]  & (UT) & [hours] & [°]  & [°] \\ \hline
1  & 2023-11-09 & 0.66 & 3    & 10  & 13:32  & 4  & -16  & 21  \\
2  & 2023-12-24 & 0.94 & 1-4  & 12  & 23:09  & 12 & -42  & 10  \\
3  & 2023-12-31 & 0.95 & 2    &  & 15:15  & 8  & -102 & 52  \\
4  & 2022-03-10 & 0.46 & 7    & 70  & 21:00  & 1  & -67  & -31 \\
5  & 2022-06-07 & 0.96 & 0.7  & 20  & 13:00  & 12 & -13  & 52  \\
6  & 2022-06-26 & 1.01 & 1    & 10  & 10:30  & 15 & 13   & 82  \\
7  & 2022-07-23 & 0.99 & 2-10 & 50  & 22:30  & 8  & -24  & 44  \\
8  & 2023-01-20 & 0.95 & 0.6  & 7   & 23:30  & 13 & -34  & 24  \\
9  & 2023-08-07 & 0.88 & 3    & 60  &       & 8  & 83   & 133 \\
10 & 2024-03-23 & 0.39 & 20   &  & 01:30  & 4  & 23   & 42  \\ \hline
\end{tabular}%
\begin{tablenotes}
    \footnotesize
        \item[I] Contaminated by a subsequent event before it ends. 
\end{tablenotes}
\end{threeparttable}
}
\caption{Summary of 10 SEP Events with IVD Features. The SEP observations by SolO and spacecraft connectivities for each event can be found in SI. }
\label{tab:results2}
\end{table}

\section*{Crossref Funding Data Registry}
The authors acknowledge the support by the National Natural Science Foundation of China (Grant Nos. 42188101, 42130204,42474221).
Solar Orbiter is a mission of international cooperation between ESA and NASA, operated by ESA. EPT and EPD are supported by the German Space Agency, DLR, under grant 50OT2002 and the Spanish MINCIN Project PID2019-104863RBI00/AEI 10.13039/501100011033.

\section*{Data Availability Statement}\label{sec:datastatement}
The authors also acknowledge the different GOES, SOHO, STEREO and SolO instrument teams, and all the data used in this study are available at the website of NASA({\url{https://cdaweb.gsfc.nasa.gov/}}).

\section*{Author Contributions Statement}
YCL has performed most data analysis and visualizations. JG has written most of the text together with YCL and DP. DP did the original observation of the phenomena, performed an initial analysis, compiled the list of events, and made some of the figures. YW proposed the DSA mechanism and initiated the theoretical analysis. All authors contributed to discussing about the methods, results and editing the text. 

\section*{Competing Interests Statement}
Authors declare no competing interests.

\clearpage
\bibliography{nsr_sample}
\bibliographystyle{unsrt}

\clearpage
\appendix
\section*{Appendix: Supplementary information}

\section{Dataset}\label{sec:dataset}

This study is based on data from several space missions observing from different longitude and radial distance from the Sun. We used in-situ and remote-sensing data from Solar Orbiter (SolO), as well as measurements from the Solar-Terrestrial Relations Observatory Ahead \citep[STEREO A,][]{Kaiser2008} and near-Earth observatories like the Geostationary Operational Environmental Satellite \citep[GOES,][]{GOESdatabook}, and the Solar and Heliospheric Observatory \citep[SOHO,][]{Domingo95}.

The unusual solar proton events are measured by the Energetic Particle Detector \citep[EPD,][]{RodriguezPacheco2020,Wimmer21} instrument suite onboard SolO, which provides comprehensive ion observations across a broad energy range. In this work, observations of vital energy ranges of IVD events are from the Electron Proton Telescope (EPT) and the High-Energy Telescope (HET) of EPD. 
EPT measures ions from 48 to 6100 keV while providing anisotropy information from four different viewing directions, which point sunward, anti-sunward, north and south. Generally for scatter-free and scatter-poor events, first particles arrive primarily from the direction of the Sun so that the sunward-looking telescope measures the earliest onset of particles (see Supplementary Information). 
Therefore, this study mainly uses the sunward-looking data for the onset analysis. 

Meanwhile, HET can discriminate between different elements and measures ions from 6.8 MeV nuc$^{-1}$ to more than 100 MeV nuc$^{-1}$, with the upper energy limit depending on the ion species. HET also provides four fields of view that share pointing directions with EPT.

SEPs are tightly associated with the eruption of solar flares which are often accompanied by CMEs and shocks \citep{reames1999}.
Observations of solar flares are carried out by the Atmospheric Imaging Assembly \citep[AIA,][]{lemen2012} onboard Solar Dynamic Observatory \citep[SDO,][]{SDOmission} in multiple EUV wavelengths and cross-checked for the higher energy ranges with the Spectrometer/Telescope for Imaging X-rays \citep[STIX,][]{krucker2020} onboard SolO when possible. 
The SDO/AIA flare observations of the 2023-11-09 event are given in the Section: Flare HXR and radio observations and Figure 1 of the main text.

Meanwhile, we characterise the associated CMEs based on coronagraph images taken by the C2 and C3 coronagraphs of the Large Angle and Spectrometric COronagraph \citep[LASCO,][]{brueckner1995} on board the SOHO spacecraft and COR1 and COR2 in the Sun Earth Connection Coronal and Heliospheric Investigation \citep[SECCHI,][]{howard2008} instrument suite provided by the STEREO-A spacecraft. 
The SOHO LASCO CME Catalogue\footnote{https://cdaw.gsfc.nasa.gov/CME\_list/} has been employed to help identify the flare and CME associated with the SEP event considering the closest eruption time to the SEP release time (explained  in Method section of the main text).
For the 2023-11-09, 2023-12-24 and 2023-12-31 SEP events, their associated flare and CME information is given in Table 1 of the main text.
Coronagraph images from both SOHO and STEREO-A are combined to reconstruct CMEs using the GCS modelling \citep{Thernisien2006} approach to obtain the initial height, timing, speed and direction of the CME (at about $\sim$ 10 solar radii). The exact parameters of the fitted CMEs are given in the Supplementary Information.

Additionally, radio observations, serving as important indicators of shock acceleration and particle propagation, are provided by the Radio and Plasma Wave Science \citep[WAVES,][]{Bougeret1995} measurements onboard WIND spacecraft, and the Radio and Plasma Waves \citep[RPW,][]{Maksimovic2020} data onboard SolO are taken as a reference to better identify the timing of the events. 
The WIND observations (which are clearer than RPW plots) are given in the Supplementary Information. The time durations of the Type II radio emission for the 2023-11-09, 2023-12-24 and 2023-12-31 SEP events are given in Table 1 of the main text.

Other instruments onboard SolO are used to understand the interplanetary plasma and magnetic environment where particles propagate through (see Figure 1(c) of the main text). Solar wind parameters such as bulk speed, proton density and proton temperature are provided by the Solar Orbiter Solar Wind Analyser \citep[SWA,][]{owen2020} suite, while the interplanetary magnetic field is measured by the Solar Orbiter magnetometer \citep[MAG,][]{horbury2020}. 

\section{Flare HXR and radio observations}\label{sec:flare_obs}

\begin{figure}
\centering
\includegraphics[width=16pc]
    {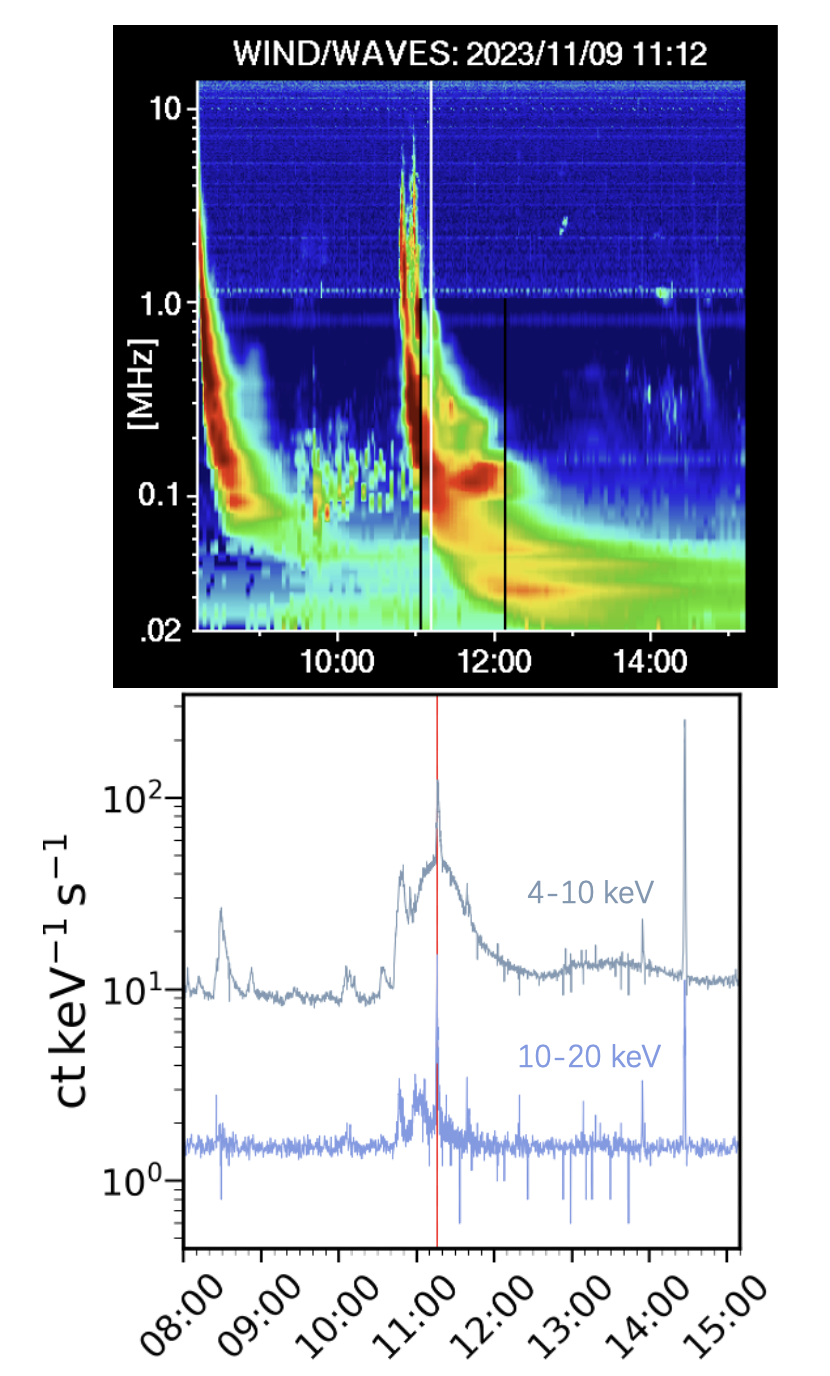}
    \centering
    \caption{(a) Type II and Type III radio burst associated with the 2023-11-09 event, observed by WIND/WAVES, with the white line indicating the Type II start time corresponding to the later flare (S10W04). (b) Soft X-ray (4-10 keV) and hard X-ray (10-20 keV) observations from SolO/STIX across two channels, with red line marking the peak of each channel. For consistency, the time axis shown in the panels represent the light arrival time at 1 AU, whereas the times mentioned in the text have been adjusted by subtracting 8.33 minutes.}
    \label{fig:1109hxr+rb}
\end{figure}
The flare closest to the VD release time (11:39 $\pm$ 7 mins) of {\bf the} 2023-11-09 event was a C2.6 class flare, lasting from 10:45 UT to 11:29 UT, peaking at 11:05 UT {\bf (time has been subtracted considering the light travel time).} Fig. \ref{fig:1109hxr+rb}(b) shows the hard and soft X-ray count rates as recorded by SolO/STIX, and the time axis is shifted to 1 AU to campare with the solar radio burst observed at 1 AU. The flare was located at W04, S10 based on SDO/AIA observations. The longitudinal separation between the flare and Solar Orbiter's magnetic footpoint as derived from the observed solar wind speed of 350 km/s was 21.3 degrees. The longitudinal separation between the flare and the Earth's magnetic footpoint as derived from the observed solar wind speed of 680 km/s was 33.3 degrees.

Although the CME speed and flare classes are not among the highest on record, strong solar radio bursts were observed.
In the observed frequency bands from WIND/WAVES,
two slow-drifting Type II radio bursts and a fast-drifting Type III radio burst at higher frequencies are presented in Fig. \ref{fig:1109hxr+rb}(a). The two Type II radio bursts correspond to two closely occurring eruptive C-class flares and their related CME shocks. The later one started at 11:04 UT spanning $\sim$1 MHz to 20 kHz. Together with the earlier occurring Type III radio burst, this strongly indicates the particle acceleration processes from the flare-CME event chain.

Table 1 in the main text shows the timing of flare X-rays and radio bursts for the 2023-11-09, 2023-12-24 and 2023-12-31 events.

\section{CME propagation derived from DBM}\label{sec:DBM}
In this study, the drag-based model \citep[DBM;][]{vrsnak2013} is employed to determine the kinematics of CME shocks, chosen for its high efficiency. The analytical solutions for acceleration, velocity and height  are given by

\begin{gather}
     a(t)=-\gamma (v(t)-v_{\text{sw}})\lvert v(t)-v_{\text{sw}} \rvert,\\
     v(t)=\frac{v_0-v_{\text{sw}}}{1+\gamma(v_0-v_{\text{sw}})t}+v_{\text{sw}},\\
     r(t)=\frac{1}{\gamma}\ln{[1+\gamma (v_0-v_{\text{sw}})t]}+v_{\text{sw}}t+r_0, 
\end{gather}

where $r_0$ is the initial radial distance, $v_0$ is the initial CME speed at $r_0$ in km/s, $v_{\text{sw}}$ is the ambient solar wind speed in km/s, and $\gamma$ is the drag parameter that determines the rate at which the CME {\bf interacts with} the solar wind. The input parameters for the four events analysed in detail are presented in Table \ref{tab:dbm}. 

\begin{table*}[]
\centering
\resizebox{\textwidth}{!}{
\begin{tabular}{ccccccccc}
\toprule
\textbf{No.} & \textbf{Date} & \textbf{Initial time} & \textbf{$r_0$ ($R_\odot$)} & \textbf{$\phi\,(\si{\degree})$} & \textbf{$\theta\,(\si{\degree})$} & \textbf{$v_0$ (km/s)} & \textbf{$v_{\text{sw}}$ (km/s)} & \textbf{$\gamma$ ($\cdot 10^{-7}$ km$^{-1}$)} \\
\midrule
\textbf{1} & 2023-11-09 & 14:57 & $17\pm 1.7$ & $6$ & $17$ & $870\pm 87$ & $350\pm 70$ & $0.1\mp 0.05$ \\
\textbf{2} & 2023-12-24 & 16:54 & $17\pm 1.7$ & $340$ & $79$ & $740\pm 74$ & $300\pm 60$ & $0.1\mp 0.05$ \\
\textbf{3} & 2023-12-31 & 13:54 & $17\pm 1.7$ & $101$ & $36$ & $830\pm 83$ & $400\pm 80$ & $0.1\mp 0.05$ \\
\bottomrule
\end{tabular}}
    \caption{DBM input parameters for the 3 events analysed in detail. $\gamma$ values between 0.1--0.2$\times$10$^{-7}$km$^{-1}$ are found to match best for simulating the CME shock-sheath structure \citep[see][]{vrsnak2014}. In this study, the uncertainties for $r_0$,$v_0$ are 10\%, for $v_{\text{sw}}$ 20\% and for $\gamma$ 50\%.}
    \label{tab:dbm}
\end{table*}

\begin{figure*}
\resizebox{\textwidth}{!}
    {\includegraphics{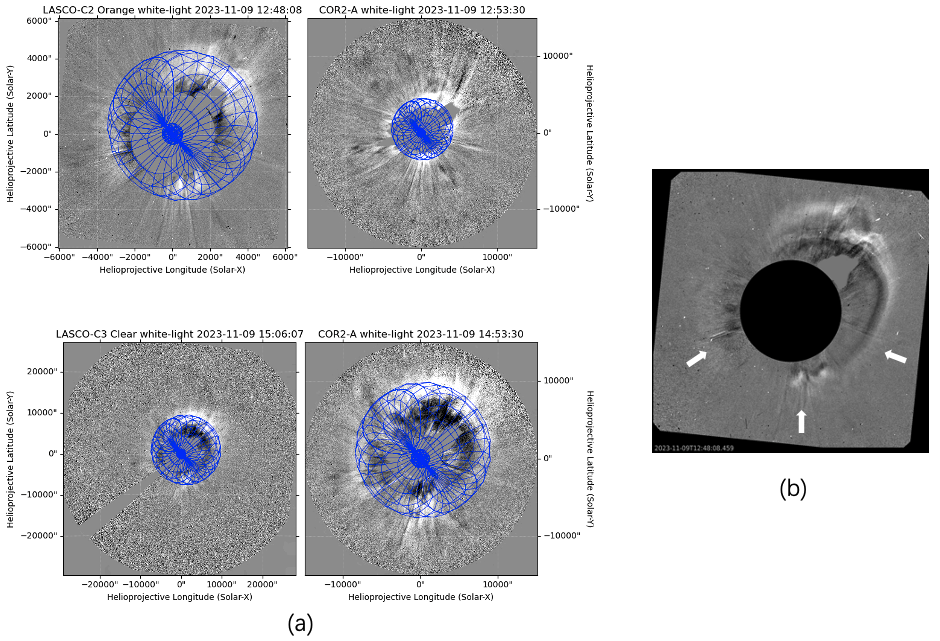}}  
    \caption{(a) GCS fitting for the shock (blue mesh) using simultaneous running difference white-light image from COR2-A/STEREO-A and LASCO/SoHO at $\sim$12:50 (first row) and $\sim$15:00 (second row) on 2023-11-09. (b) Running difference white-light image from LASCO C2/SoHO at 12:48. Note that another flare, occurring slightly before the SEP-related flare in the northern hemisphere, is associated with a northward CME. This CME interacted with a subsequent southward halo CME, that is related to the SEP event and marked by three arrows at the image's edge.   }
     \label{fig:1109gcs}
\end{figure*}
Inputs related to shock geometry such as $r_0$ and initial time, are derived from the Graduated Cylindrical Shell (GCS) modelling \citep[see][]{Thernisien2006}. To re-construct the 3D geometry, propagation direction and kinematics, GCS requires white-light observations of a CME from at least two viewpoints. The software developed by Forstner \citep{forstner} used in this study, adapts combinations of COR2-A on STEREO-A and LASCO C2/C3 on SoHO. For shock simulation, the fitting can be accomplished by setting the half-angle to zero and the aspect ratio close to one \citep{temmer2015interplanetary}.  

Although DBM is used to describe the kinematic evolution of CME magnetic structures, Vr{\v{s}}nak et al. \citep{vrsnak2014} found a parameter range of $\gamma$ values for which the arrival times of shocks aligned well with the results from ENLIL shock front tracking \citep[see also][]{Dumbovic2021}. Studies from e.g., Dumbovic et al.2018, Temmer et al.2015, Guo et al. 2018 \citep{dumbovic2018,temmer2015interplanetary,guo2018}, have all applied the DBM in CME shock simulations with different empirical $\gamma$ values in different situations. Here in our work, we adapt $\gamma=0.1 \cdot 10^{-7} \text{km}^{-1}$ which is consistent with the values used to compare DBM and ENLIL shock simulation in Vr{\v{s}}nak et al. 2014 \citep{vrsnak2014}.

Using GCS fittings on the early stage shock, the shock speed $v_0$ can be determined from quasi-GCS measurements \citep{temmer2015interplanetary}. Figure \ref{fig:1109gcs}(a) shows examples of shock fitting for the 2023-11-09 event at two time points, derived from running-difference white-light images. These fittings are utilized to estimate the shock speed at the respective times.

Due to the difficulty of measuring ambient solar wind speed $v_{\text{sw}}$ throughout the CME propagation, various methods have been employed to determine the value of $v_{\text{sw}}$ when using DBM. In our study, for IVD candidates linked to a CME that has passed SolO, we directly use solar wind speed measurements of the shock {\bf upstream} from SWA measurements. For those not impacted by the CME shock at SolO, we use inferred values from the DBEMv4 solar wind module\footnote{https://swe.ssa.esa.int/graz-dbem-federated} which have significant uncertainties. So both the modelled and measured solar wind speeds have been assigned a 20\% uncertainty, as shown in Table \ref{tab:dbm}. 

\begin{figure}
\centering
\includegraphics[width=16pc]{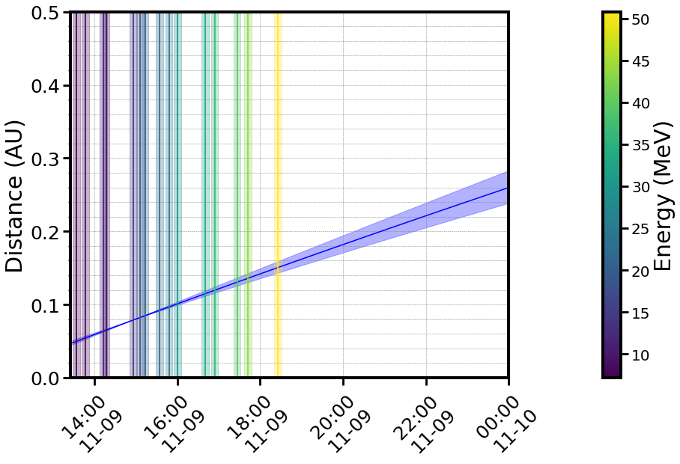}  
    \caption{The derived release time and distance for IVD particles and the propagation time and distance of the DBM-modelled shock for the 2023-11-09 event. The blue line indicates distance, while the blue shadow band represents the output uncertainty due to the input uncertainties mentioned earlier (listed in Table 1). Vertical lines marked the release time $t_{\text{release}}(E^{'})$ of IVD particles at different energies, indicated by the colour bar on the right. The uncertainty of $t_{\text{release}}(E^{'})$ is calculated using error propagation methods accounting for the observational time interval and energy range.}
    \label{fig:1109dbm}
\end{figure}

By applying the DBM, the energy-dependent release phase of IVD particles in 2023-11-09 event within the CME propagation can be obtained, as shown in Fig. \ref{fig:1109dbm}. 
The IVD particles, ranging from 11 to 53 MeV, were released over a duration lasting more than 4 hours. These particles are released sequentially from lowest to highest energy, when the shock was below 0.16 AU. This feature of sequential release according to energy is evidence of the diffusive acceleration mentioned in the main text, as well as an explanation for the IVD phenomenon.

\section{Directionality of SEPs during the 2023-11-09 event}\label{sec:anisotropy}

\begin{figure*}
\resizebox{\textwidth}{!}
    {\includegraphics{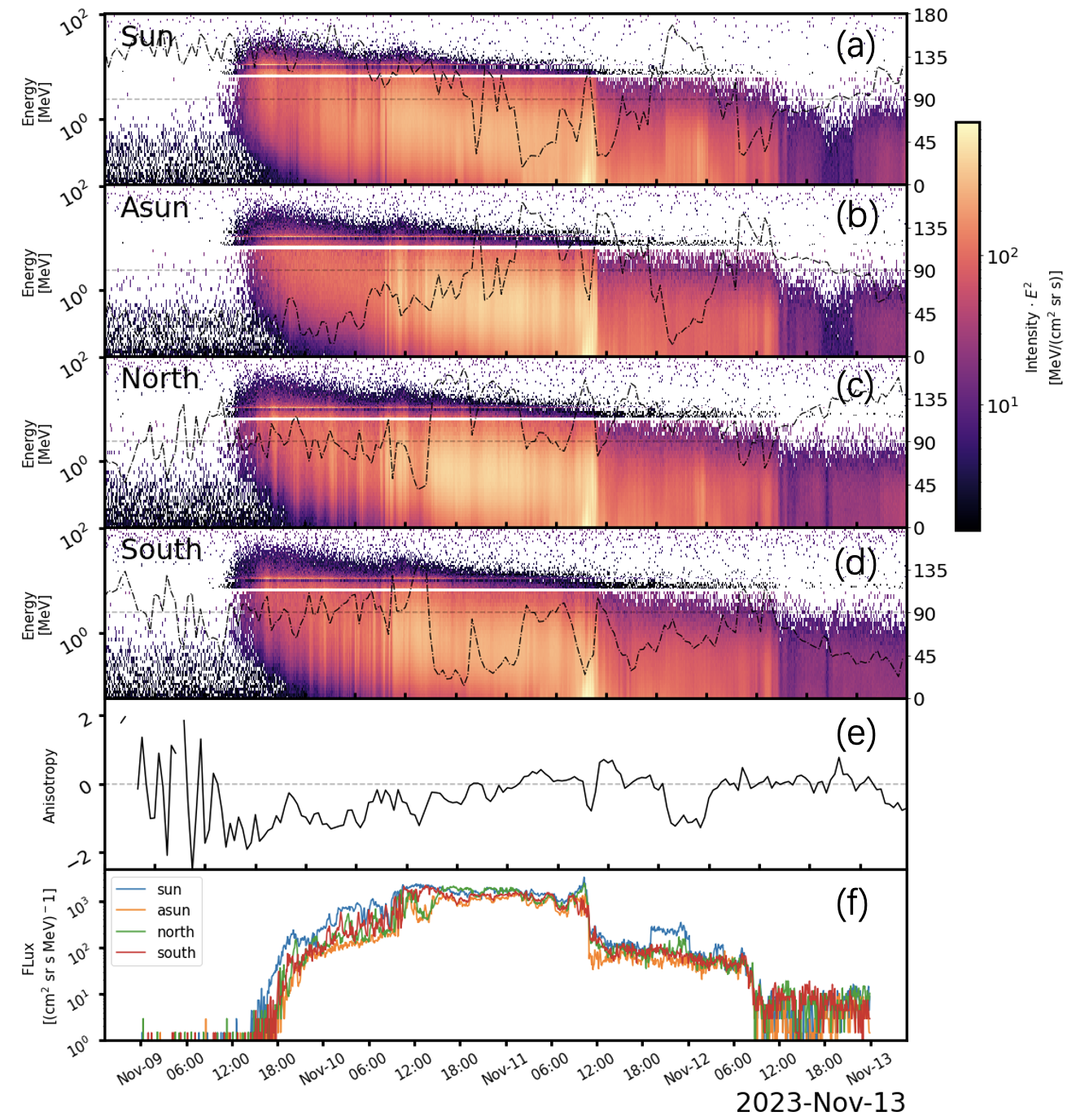}}  
    \caption{Top 4 panels are EPT observations from different telescopes for protons, as indicated in the upper-left corner of each panel. Pitch angle is also shown as a dot-dashed line and is represented by the scale on the right-hand axis. (e) shows anisotropy of the first arrival energy, $\sim$1 MeV, while (f) shows the intensity-time profile of the same energy channel.}
    \label{fig:1109an}
\end{figure*}

Particle directionality at the onset of the SEP events is shown in Fig. \ref{fig:1109an} with the 2023-11-09 event as an example based on measurements from the four apertures of the EPT and HET. The top four panels display the intensity colour coded multiplied by $E^2$ for the combined EPT and HET measurements of the four fields of view (i.e., Sun, anti-Sun, north, and south). The pitch angle at the centre of each aperture is indicated by a dash-dotted curve within each panel, scaled to the right-hand y-axis. A faint dash-dotted line marks 90 degrees as a reference.
The earliest and most distinct IVD features were noted in the sunward-facing EPT and HET at approximately 15:00 UT on 9 November, with similar observations following about an hour later in the south-facing telescope. In contrast, such phenomena in the anti-sunward and north orientations were only detected in the HET energy channels. 

First-order anisotropy A is defined as the following equation:

\begin{gather}
    A=\frac{3\Sigma_{i}^{}I(\mu _i)\cdot\mu_i \cdot \delta \mu_i}{\Sigma_{i}^{}I(\mu_i)\cdot \delta\mu_i}
\end{gather}
where $I(\mu)$ is the pitch-angle-dependent intensity measured by the $i$ viewing direction and $\mu$ is the average pitch angle cosine of the direction \citep{Dresing2014}. 
The coverage of the pitch angle depends on the orientation of the magnetic field relative to the aperture of the telescopes. EPT and HET apertures do not cover the entire sky area and hence do not observe the complete $\mu$-space. Consequently, the reconstruction of the pitch-angle distribution is limited by the four directions of the observations. 


The last two panels present strong anisotropy of the $\sim$1 MeV proton at the onset stage around 15:00 UT, November 9. Panel (f) shows the intensity-time profile for the 1 MeV protons from different EPT telescopes. The sunward telescope observed not only the earliest onset but also the highest flux. In panel (e), $A$ in the onset period shows a large absolute value ($\lvert A\rvert\approx2$) and lasts for approximately 6 hours. Coupled with the earliest onset of sunward particles shown in panel (f), this suggests a significant anisotropy in the arrival times of particles from different orientations. The anisotropy is negative, meanwhile the $B_r$ shown in (see Fig. 1(c) in main text) is negative. This indicates that the initial acceleration site of these particles was at the Sun. 
{\bf As the event continues, the absolute value of anisotropy converges towards zero except for a short duration of fine structure from 19:00 of Nov 11 until 00:00 of Nov 12. This is known as the reservoir effect \citep{reservoir1992} during the later phase of SEP events due to strong particle scatter process in the heliosphere.}



\section{SEP observations of 2023-12-24 and 2023-12-31 events}\label{sec:10events}

\begin{figure*}
\resizebox{\textwidth}{!}
    {\includegraphics{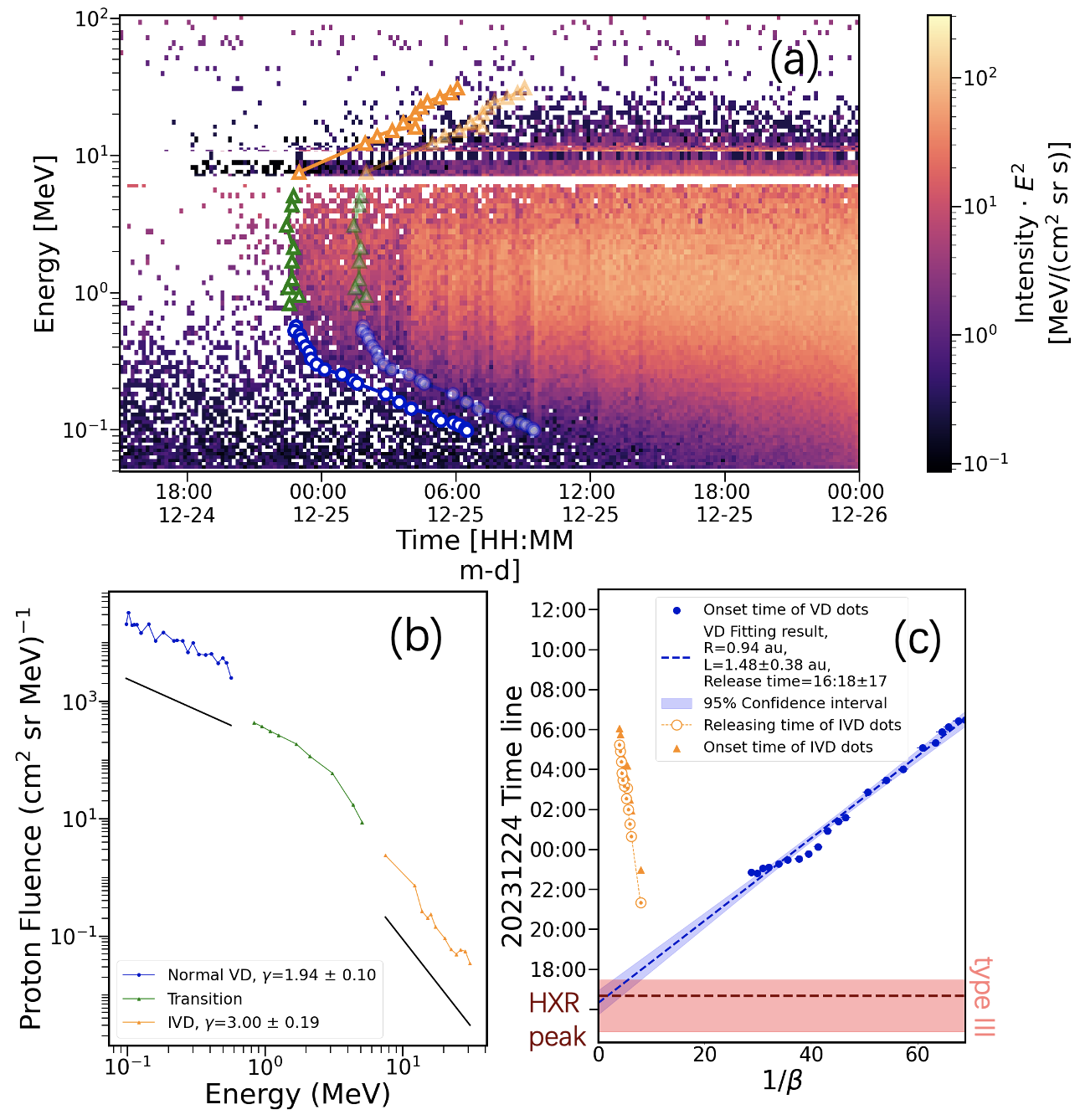}} 
     \caption{(a) Dynamic spectra of the early phase of the 2023-12-24 event; (b) The release time and path length result from VDA of protons under $\sim$0.7 MeV and from IVD release time analysis of protons above $\sim$ 7 MeV and (c) The 3-hour-integrated energy spectra of proton measurements, starting from the onset of each energy range.
    In (a), the onset time for VD particles is marked with blue circles, while the onset time for IVD particles is indicated with orange triangles (for onset time determination see Method in the main text). The first arrival energy within the transition range between VD and IVD energy is marked with green triangles.
    The onset time is used in panel (b) following VDA method (see Method in main text) to derive the release time and path length of energetic protons shown in the legend. 
    Timing of the M2.6 flare hard X-ray ($\sim$ 10 keV) and related radio bursts shifted to the solar surface (subtracting 8 min accounting for the time of photons reaching SolO at 0.94 au) is marked by horizontal dashed lines and pink band, respectively. In (c), the power-law fitting obtained for the low- and high-energy parts is marked by the black lines and the legends. 
     }
    \label{fig:1224vda}
\end{figure*}

Figure \ref{fig:1224vda} shows the measurements and VDA/IVDA results for the 2023-12-24 event. In panel (a) we depict the dynamic spectra with the proton measurements from EPT and HET. Blue circles mark the onset time for each energy bin in the lower energy range, corresponding to the regular VD pattern, and the orange dots show the onset for the energy range in the IVD part. All onsets were selected manually. In panel (b) we plot the 3-hour-integrated spectra for both energy ranges starting from the onset of each energy range, using the same colours for VD and IVD as in panel (a). 
This event was observed by SolO from a radial distance to the Sun of 0.94 au and under a measured solar wind speed of 460 km/s. The results of the VDA in panel (c) suggest a release time of 16:18 UT $\pm$ 17 min, and a {\bf travelled} path length of 1.48 $\pm$ 0.38 au, comparable with the 1.04 au calculated for the correspondent Parker spiral.

\begin{figure*}
\resizebox{\textwidth}{!}
    {\includegraphics{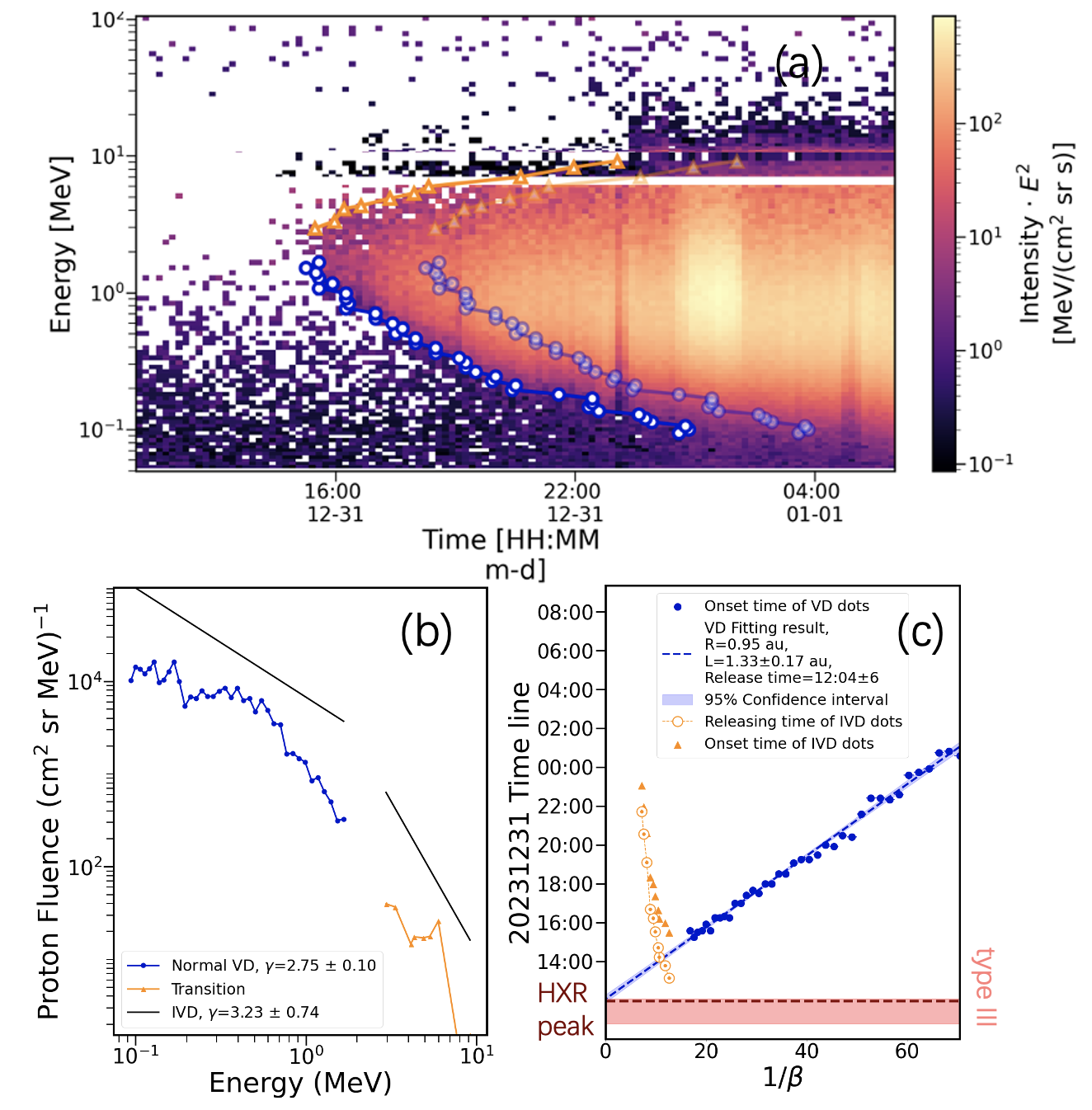}}  

    \caption{Same as Fig.~\ref{fig:1224vda} for the event on 2023-12-31.
     }
    \label{fig:1231vda}
\end{figure*}

Similar to the previous figure, Figure \ref{fig:1231vda} presents the measurements and VDA/IVDA results for the 2023-12-31 event. This event was observed by SolO at a radial distance to the Sun of 0.95 au and under a measured solar wind speed of 393 km/s. The VDA results suggest a release time of 12:04 UT $\pm$ 6 min, and a travelled path length of 1.33 $\pm$ 0.17 au, comparable with the 1.09 au calculated for the correspondent Parker spiral.

\section{Information of other IVD events}\label{sec:10vda}

\begin{figure*}
\resizebox{\textwidth}{!}
    {\includegraphics{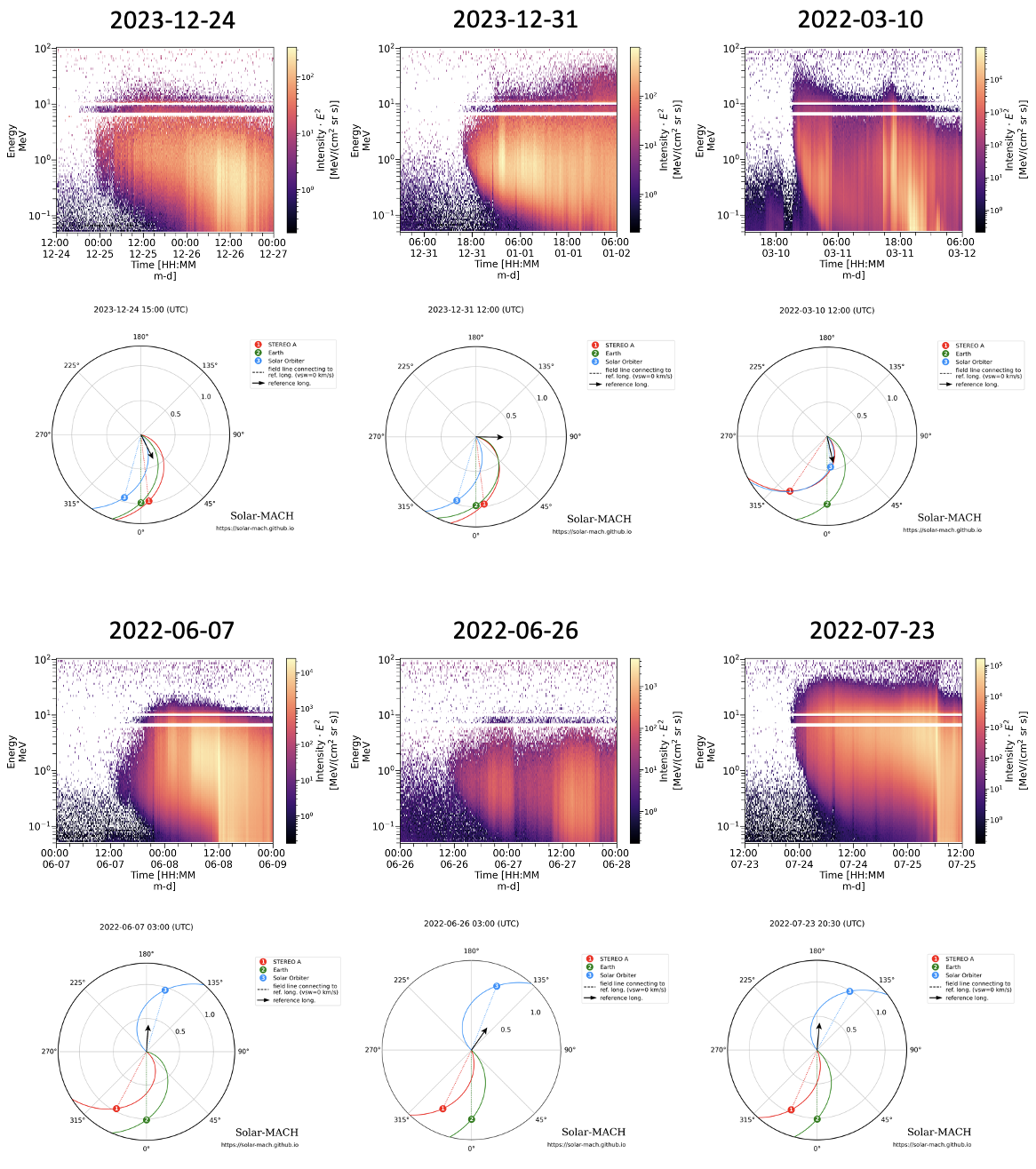}}  
    \caption{Dynamic spectra of 6 of the 10 IVD events listed in Table 2 of the main text, along with the configuration plot with the location of solar source. In each longitudinal configuration, the red, green, and blue markers and curves represent STEREO A, Earth, and SolO along with their respective Parker spirals. The black arrow indicates the location of the flare most likely associated with the SEP event.
     }
    \label{fig:9event1}
\end{figure*}

\begin{figure*}
\resizebox{\textwidth}{!}
    {\includegraphics{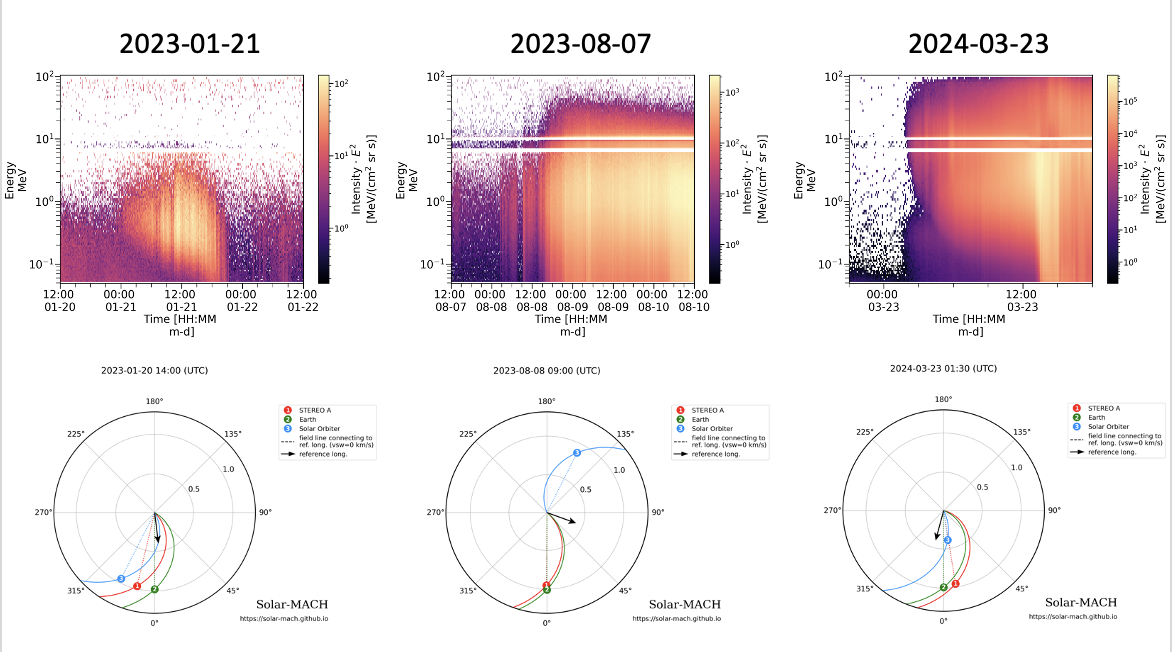}}  
    \caption{Same as Fig.~\ref{fig:9event1} for the other 3 events listed in Table 2 of the main text, except for the 2023-11-09 event discussed in detail.}
    \label{fig:9event2}
\end{figure*}

As a reference, Fig. \ref{fig:9event1} and Fig. \ref{fig:9event2} show the 9 IVD events mentioned in Table 2 of the main text and location of the identified flare related to each of them. The release time derived through VDA, combined with EUV images from The Extreme Ultraviolet Imager (EUI) onboard SolO and white-light observations from SOHO's C2, helps determine the related eruptions.
There are a few noteworthy points:
\begin{itemize}
    \item 2022-06-07 event: the onset of VD particles is divided into two segments, each showing a strong linear relationship, leading to different release times. The release time preceded the eruption time adapted in Table 2 by two and a half hours, and this eruption was the most intense flare within this time period.
    \item 2022-06-26 event: Prior to the inferred release time, there were a series of eruptive events, including two flares with close longitudes in the southern and northern hemispheres, respectively, and a filament eruption near the eastern limb (from SolO's FOV). The configuration in Table 2 shows the direction of the northern flare, which was related to a CME.
    \item 2023-08-07 event: this event shows an unusual and complex onset phase that could not be determined using CUSUM or manual selection to identify the onset time of the VD part. Additionally, EUI observations were missing during the event, and according to X-ray observation from STIX, multiple eruptions occurred on the day of the SEP onset, plus an X-class flare the day before. The eruption directions plotted in Fig. \ref{fig:9event2} (and recorded in Table 2 of the main text) is derived from the flare location tool on the STIX quick-look website\footnote{ \url{https://datacenter.stix.i4ds.net/}}, which indicates the concentrated positions of eruptions.
\end{itemize}

\end{document}